\documentclass[aps, prd, floatfix, nofootinbib, superscriptaddress, twocolumn]{revtex4-1}

\usepackage[utf8]{inputenc}
\usepackage[brazil]{babel}
\usepackage[T1]{fontenc}
\usepackage{latexsym}
\usepackage{amsmath}
\usepackage{amssymb}
\usepackage{amsfonts}
\usepackage[mathscr,scaled=1.15]{urwchancal}
\DeclareFontFamily{OT1}{pzc}{}
\DeclareFontShape{OT1}{pzc}{m}{it}%
{<-> s * [1.15] pzcmi7t}{}
\DeclareMathAlphabet{\mathpzc}{OT1}{pzc}{m}{it}

\usepackage{color}

\usepackage{supertabular}
\usepackage{placeins}
\usepackage{epsfig}
\usepackage{graphicx}

\definecolor{purple}{rgb}{0.5,0,0.5}
\definecolor{blue}{rgb}{0.0,0,0.9}
\definecolor{prdblue}{rgb}{0.133,0.118,0.498}
\usepackage[colorlinks=true, pdfstartview=FitV, linkcolor=prdblue, citecolor= prdblue, urlcolor=prdblue]{hyperref}

\begin{document}

\title{Estudo das ondas gravitacionais no ensino médio: uma abordagem teórica e experimental \\ {\small Study of gravitational waves in high school: a theoretical and experimental approach}}

\author{Brendon S. M. Barros}
\email{brendonmbmobile@gmail.com}
\affiliation{\mbox{Colégio São Lucas}, Picos, PI, Brasil}
\affiliation{\mbox{Instituto Federal de Educação, Ciência e Tecnologia do Piauí, Grupo de Pesquisa em Ensino de Física},\\ Picos, PI, Brasil}

\author{Emanuel V. de Souza} 
\email{emanuel.veras@ifpi.edu.br}
\affiliation{\mbox{Instituto Federal de Educação, Ciência e Tecnologia do Piauí, Grupo de Pesquisa em Ensino de Física},\\ Picos, PI, Brasil}


\date{03 Fevereiro 2020}

\begin{abstract}
Neste trabalho apresentamos de forma teórica, histórica e prática a inclusão das ondas gravitacionais no ensino de física. Através da explanação do tema por via de debates e de práticas experimentais em sala de aula, discutimos a presença da Física Moderna e Contemporânea (FMC) no Ensino Médio, apresentando pontos positivos e mostrando acima de tudo que os alunos possuem a capacidade de assimilar conceitos físicos modernos e relacioná-los com o seu cotidiano e discutindo também as barreiras para essa inclusão no ensino de física. Além da apresentação teórica introdutória das ondas gravitacionais, o trabalho que se segue também levanta a importante questão da prática experimental em sala de aula, com a construção de um interferômetro de Michelson-Morley de baixo custo, usando-o como um comparativo ao detector LIGO (Laser Interferometer for Gravitational Waves Observatory) e embasando mais ainda a comprovação das ondas gravitacionais para os alunos em sala. Com toda essa construção teórica e experimental das ondas gravitacionais, o trabalho apresenta indicadores e dados fundamentais para essa inclusão com o objetivo de aperfeiçoar o ensino de física na educação básica.\\
\textbf{Palavras-chaves}: Ondas Gravitacionais; Ensino de Física; Física Moderna e Contemporânea.  \\

In this paper we present in a theoretical, historical and practical way the inclusion of gravitational waves in the teaching of physics. Through the explanation of the theme through debates and experimental practices in the classroom, we discuss the presence of Modern and Contemporary Physics (MCP) in High School, presenting positive points and showing above all that students have the ability to assimilate modern physical concepts and relate them with their day by day and also to discuss the barriers to this inclusion in teaching of physics. In addition to the introductory theoretical presentation of gravitational waves, the following work also raises the important question of classroom experimental practice, building a low-cost Michelson-Morley interferometer, using it as a comparison to the LIGO (Laser Interferometer for Gravitational Waves Observatory) detector detector and basing even more so is the evidence of gravitational waves for students in class. With all this theoretical and experimental construction of gravitational waves, the work presents indicators and fundamental data for this inclusion in order to improve the teaching of physics in basic education. 
\end{abstract}

\maketitle


\section{Introdução}

O século XXI tem sido um grande momento para o desenvolvimento da física e da ciência como um todo. Grandes passos foram dados e grandes revoluções nas ideias sobre o funcionamento do Universo foram geradas e/ou reafirmadas, trazendo a sensação de que a física se mantém inovadora a ponto de ainda modificar nossas concepções sobre o conhecimento do Universo.

O final do ano de 2015 e início de 2016 trouxe consigo essa sensação de trabalho contínuo da física, em comprovar, algo predito por Albert Einstein há mais de 100 anos, 1916, a existência das Ondas Gravitacionais (OG) pelo LIGO (Laser Interferometer for Gravitational Waves Observatory)~\cite{ligo} em colaboração com o Virgo Interferometer~\cite{virgo} e outros detectores. A detecção foi possível com a ocorrência da fusão de um sistema Binário de Buracos Negros (BBN's), como abordado por Cattani e Bassalo~\cite{Cattani} em afirmar que BBN's podem ser um fonte potencial para essas ondas e um meio pelo qual poderiam ser comprovadas. Essa detecção certamente marca uma nova fase da astronomia, a chamada \textit{Astronomia de Ondas Gravitacionais}.

A formulação da existência de OG e principalmente da sua elaboração como teoria passou por um processo árduo, principalmente devido as interrupções sofridas pela eclosão tanto da 1º Guerra mundial (1914-1918), quanto da 2ª Guerra mundial (1939-1945) e seu respectivo pós-guerra, fazendo com que boa parte das pesquisas da época entrassem em modo \textit{stand by}, afim de que praticamente todo e qualquer financiamento, desenvolvimento científico e dedicação de cientistas fossem direcionados a guerra, resultando em que muitas dessas pesquisas entrassem no total esquecimento. Um caso em particular que praticamente entrou em parcial esquecimento foi justamente a teoria da Relatividade Geral (RG) de Einstein, por ser, na época, uma teoria especulativa, sem uma formulação matemática convincente e um pouco vazia de conteúdo físico, como menciona Saa~\cite{Saa}. No entanto, devido à grande indagação para a comunidade científica, sobre a existência das ondas gravitacionais, a RG, posteriormente, foi retomada.

Além da imposição das guerras nos países europeus, muitas discordâncias, até do próprio Albert Einstein, ainda persistiam, possuindo em contrapartida muitas aprovações e crenças de sua existência, envolvendo uma gama de físicos empenhados em trazer para comunidade científica algo tão vislumbrado pela sua ideia inicial de unificação e resolução de problemas que a mecânica newtoniana não era capaz de resolver. Esse é o embate característico do desenvolvimento das OG citado por Kennefick~\cite{Kennefick}.

Sendo uma previsão resultante da Relatividade Geral, como afirma Saa~\cite{Saa} em dizer que, "esta deve ser uma previsão de qualquer teoria que rejeite interações instantâneas à distância, como as presentes na Gravitação Universal de Newton", ainda não estava concretizada e aceita por boa parte da comunidade científica. Muitos foram os incrédulos à teoria das ondas gravitacionais e principalmente a sua comprovação experimental, mas aparentemente essa incredulidade e alto nível de ceticismo foi o combustível primordial para fazer com que a sua comprovação saísse de apenas especulações para fatos incontestáveis. 

A formulação da teoria das OG, mesmo sendo proposta em 1916 por Einstein, só foi começar a encontrar caminhos fortificados para sua estruturação nos anos de 1950 e mesmo assim muitos anos à frente de intensas discussões e debates até então dar início, verdadeiramente, da tentativa de detecção dessas ondas.

 O próprio LIGO demorou décadas para ter a sua primeira comprovação, 14 setembro de 2015, a chamada GW150914 e no mesmo ano, em 26 de dezembro, a segunda comprovação, chamada de GW151226, ambas em interferômetros de mesma base de funcionamento, uma situada em Hanford, Washington (EUA), e a outra em Livingston, Louisiana (EUA), para então chegar a impressa mundial e divulgar seus dados e comprovações, impactando de forma positiva e estimulante a sociedade científica, trazendo novas portas de pesquisas e campos de conhecimento como a astronomia gravitacional.

Como mencionado anteriormente, essas descobertas e comprovações trazem a sensação vívida do trabalho da física na continuidade de modificar o pensar humano, mas em contrapartida o seu ensino não desperta essa sensação nos discentes, por se manter preso e estagnado a conceitos ultrapassados e antigos da ideia de funcionamento da física e do Universo, como menciona Moreira~\cite{Moreira}, 
" (...) o ensino da Física estimula a aprendizagem mecânica de conteúdos desatualizados. Estamos no século XXI, mas a Física ensinada não passa do século XIX".

Sem quase nenhum espaço para apresentação de conceitos novos, ideias novas e concepção da Física Moderna e Contemporânea em sala de aula, a física tende a ser direcionada de forma errônea, gerando uma concepção desvirtuada da ideia de conhecimento do discente que tem contato com o ensino de física. Temos conhecimento que essa deficiência da presença do ensino da Física Moderna e Contemporânea nas escolas, não se dá apenas por um único fator, mas sim, por uma série de fatores e variáveis que em conjunto auxiliam na desconstrução do pensamento físico e principalmente científico dos discentes, sendo assim, o ponto de discussão deste trabalho.

Pensando nessa problemática, o respectivo trabalho tem como objetivo fundamental levar para sala de aula a ideia das ondas gravitacionais e o seu impacto perante o ensino, quebrando e desmistificando o pensamento de que alunos, "leigos à ciência", não possuem capacidade de absorver tais ideias e que esse é um tipo de conhecimento destinado apenas para um grupo seleto de cientistas. Também discutimos a relação entre a descoberta das ondas gravitacionais e assuntos que eles (discentes) veem em sala de aula, como fundamentos de óptica e gravitação, que servem como uma porta de entrada desse conteúdo em sala de aula, como também gerar conceitos novos sobre o funcionamento do Universo, mostrando a importância da renovação do conhecimento e escolhendo formas didáticas de aplicá-las como explica Moreira~\cite{Moreira} sobre a interação cognitiva entre conhecimentos novos e prévios como chave da aprendizagem significativa.

Outro enfoque que o respectivo trabalho traz é confirmar a importância da experimentação em sala de aula que surge como “(...) uma forma do aluno entrar em contato com a realidade, com a intenção de comprovar modelos ou teorias, ou ainda com o objetivo de motivar o aluno e despertar seu interesse pelo tema"~\cite{Silva}. Visando isso, além de ter a proposta de levar para a sala de aula a ideia da fundamentação teórica das ondas gravitacionais, também tem como enfoque mostrar o modelo experimental do interferômetro de Michelson-Morley, construído com materiais de baixo custo, que traz a mesma concepção de funcionamento que o LIGO utilizou para a detecção das OG, fazendo com que o aluno elucide suas ideias sobre o tema e entenda que conceitos físicos não são meras ideias prontas e matematizadas, mas sim que passam por um processo cientifico robusto e uma experimentação aguçada  a fim de trazer a verdade sobre o funcionamento da natureza. 

Para uma melhor apresentação, este trabalho está organizado da forma como segue. Na seção 2, apresentaremos uma breve discussão sobre a teoria da Relatividade de Einstein e o desenvolvimento histórico e científico das ondas gravitacionais como teoria e consequentemente o seu achado e dada comprovação. Após uma apresentação concisa do objeto deste trabalho, na seção 3 discutiremos a aplicação do conceito de ondas gravitacionais dentro da sala de aula, com o auxílio do experimento de Michelson-Morley, onde coletamos dados e discutimos ideias fundamentais ao desenvolvimento do ensino de física apresentados na seção 4. Por fim, na seção 5 apresentaremos nossas considerações finais.           


\section{Apresentação histórica, conceitual e experimental das ondas gravitacionais}

Nessa seção iremos abordar de forma concisa os principais acontecimentos históricos que contribuíram para o desenvolvimento teórico e experimental das ondas gravitacionais. Após essa abordagem, discutiremos sobre as possibilidades de inserção da temática para o ensino médio.

\subsection{Noções sobre a teoria da relatividade de Einstein}

Em 1905, ano conhecido como \textit{annus mirabilis}~\footnote{Expressão do latim que significa \textit{ano miraculoso}.} da física, Einstein traz ao mundo novas concepções sobre a ideia de movimento que até então tinha se estagnado como teoria pronta e completa, advindo de cientistas importantes como Galileu Galilei e Sir Isaac Newton. As leis fundamentais da mecânica pareciam funcionar de forma coerente a realidade em qualquer aspecto, tratando o estudo dos movimentos como algo natural à condição humana. Com o advento do desenvolvimento da física, principalmente na área do eletromagnetismo, com respeito à constante da luz em 1865 por James Clerck Maxwell, muitos conceitos relacionados ao movimento tiveram que mudar ou serem revistos.

Surgiu então a conhecida questão de Einstein que foi responsável por umas das primeiras e grandiosas formulações físicas do mesmo, em se questionar: Se pudéssemos então viajar em um feixe de luz, como observaríamos um raio de luz? Essa indagação paradoxal levantou de forma icônica o questionamento de Einstein. Como resposta, forneceu a seguinte afirmação: veríamos esse feixe de luz imóvel e sem alterações, como algo em repouso. A resposta  parecia um tanto absurda, pois sabia-se na época que um feixe de luz não era imóvel e nunca estaria em repouso. Einstein parecia um pouco duvidoso do seu pensamento, mas atribuiu que se trabalhasse com a ideia de referencial talvez essa resposta duvidosa não parecesse tão perturbadora. Entra então, na física, um dos primeiros momentos em que o termo referencial apareceria como algo decisivo. Einstein gerou outro questionamento, agora sobre as características de movimento em diferentes referenciais inerciais, questionando alguns parâmetros que até então eram considerados absolutos. Einstein então teve um artigo publicado na revista alemã, \textit{Annalen der Physik}, intitulado  \textit{“Sobre a eletrodinâmica dos corpos em movimento”}, em 1905, com base em trabalhos de Lorentz e Poincaré, trazendo consigo o fundamento da chamada teoria da Relatividade Restrita ou Especial.

O pensamento de Einstein não era algo tão novo e revolucionário como parecia ser. Galileu Galilei já se questionava a presença ou não de referenciais para então se fundamentar um movimento, quando formalizava a teoria inercial, mencionando que de fato a ideia de movimento absoluto não era algo a se crer, mas que dependia incisivamente de um referencial para ser então formulado. A questão de corpos inerciais em movimento uniforme não ter a condição de percepção e distinção entre movimento e repouso sem um dado referencial já era debatida, como menciona os autores da Ref.~\cite{Porto} em dizer que “em outras palavras, Galileu incorporou o conceito de relatividade do movimento, formulando-o, em forma de princípio”. O próprio Galileu~\cite{Galileu} reafirma essa ideia também por um trecho de seu livro  \textit{"Diálogo sobre os dois máximos sistemas do mundo ptolomaico e copernicano"}, na pele do personagem Salviati afirmando que “o movimento é movimento e como movimento opera, enquanto tem relação com coisas que carecem dele; mas entre as coisas que participam todas igualmente dele, nada opera e é como se ele não fosse”.

Tais pensamentos influenciariam de forma direta a formulação da Relatividade Especial para referenciais inerciais, gerando dois postulados fundamentais: 

\begin{enumerate}
\item As leis da Física são as mesmas em todos os referenciais inerciais; 
\item A velocidade da luz no vácuo tem o mesmo valor de $c$ em qualquer referencial inercial, independente da velocidade da fonte de luz.
\end{enumerate}

O primeiro postulado tem como principal característica trazer à física o caráter invariante, formulando então uma ideia de teoria unificada para a concepção física. O segundo postulado foi o mais robusto conceito capaz de dar fim, de uma vez por todas, à ideia de \textit{"éter luminífero"}~\footnote{A ideia de éter luminífero surgiu, no Séc. XIX, da necessidade de algo que pudesse preencher o espaço dando à luz um meio na qual pudesse se propagar, da mesma forma que acontece ao som, por exemplo, que necessita de um meio material para se propagar; utilizando-se sempre de analogias iniciais para a formulação teórica da ideia física.}, que já tinha sido questionado experimentalmente por Michelson-Morley em 1887, de forma indireta, com o experimento de interferômetro de luz.

Tudo apresentado na idealização da Relatividade Especial, que não será diferente na formulação futura das ondas gravitacionais onde reafirma-se como resultado do desenvolvimento da teoria da Relatividade, possui um caráter unificador e com uma influência direta de interferômetros para a sua concretização experimental. No caso das OG dada pela LIGO, tendo como diferencial a correlação direta com a gravitação de Newton, se formula uma nova visão sobre a gravidade que não tem mais como característica primordial a ação direta e instantânea de corpos por efeitos gravitacionais como era apresentado na gravitação do século XVIII.

Claro que um dos enfoques do respectivo trabalho é dar um embasamento teórico da Relatividade Geral de Einstein, em caráter de familiarização, para fornecer ao leitor uma mínima condição de entendimento das ondas gravitacionais. Mas é quase que impossível falar de Relatividade Geral sem mencionar os avanços de pensamento trazidos pela Relatividade Especial.

Aparentemente, Einstein não estava satisfeito com a particularidade que a Relatividade Restrita trazia ao tratar o movimento apenas em referenciais inerciais, como reafirma os autores da Ref.~\cite{Peruzzo} em dizer que “Einstein ficou insatisfeito com a Relatividade Restrita. Sentia uma forte necessidade de generalizar o princípio da relatividade dos movimentos uniformes a todos os movimentos”. Einstein era um grande defensor da ideia de unificação, sendo isso já mostrado no desenvolvimento da Relatividade Especial, e então vivificado mais ainda na formalização da Relatividade Geral, em 1916, publicando na revista Alemã, \textit{Annalen der Physik}, um artigo de 60 páginas com o título \textit{“Fundamentos da Relatividade Geral”}, que segundo o próprio Einstein, “a teoria da Relatividade Especial, comparada com a Relatividade Geral, é brincadeira de criança” e Videira~\cite{Videira} relatando que “a teoria da Relatividade Geral é uma das mais profundas, deslumbrantes e belas produções do espírito humano de todos os tempos”.

Einstein trazia consigo toda uma incorporação matemática e idealista sobre a generalização da Relatividade Especial, tornando-a mais completa e abrangente, tentando responder a dois princípios fundamentais que aqui serão apresentados:

\begin{enumerate}
\item A compatibilidade com a teoria gravitacional, após o avanço das leis de Maxwell;
\item Sobre a movimentação de corpos que não eram considerados referenciais inerciais, a fim de trazer uma covariância definitiva no entendimento da dinâmica dos corpos.
\end{enumerate}

Esses dois questionamentos fundamentaram a ideia geral, gerando novas conclusões de como o Universo funciona segundo a Relatividade.

Como já mencionado na seção anterior, a mecânica newtoniana teve seu sucesso concedido tanto pela sua veracidade, quanto por fazer correlações com a gravitação e prescrever movimentos tanto de corpos na Terra como fora dela, trazendo um avanço científico inigualável para a época. Da mesma forma, a Relatividade Especial que mencionava sobre corpos que viajavam próximo ou na velocidade da luz, sofrendo assim os seus efeitos e gerando uma nova mecânica para referenciais inerciais, era preciso também alargar essa fronteira e se compatibilizar com a gravitação, como reafirma os autores da Ref.~\cite{Peruzzo} em dizer que “a generalização da teoria da Relatividade Especial surgiu da necessidade de compatibilizá-la com a descrição da gravidade”. E por que compatibilizar com a gravitação? Qual a necessidade? As respostas dessas perguntas vêm da própria influência do avanço que Maxwell trouxe, implementando de forma aberta o termo \textit{campo}, saindo da visão newtoniana de força como algo localizado e instantâneo.

Os corpos sobre efeitos gravitacionais não seriam mediados por linhas retas e forças localizadas, mas sim como interação de campos gravitacionais. Os autores da ref.~\cite{Peruzzo} explicam um pouco sobre essa interação de campos mostrando intuitivamente que,

\begin{quote}
Quando se larga um objeto nas proximidades da Terra ele cai em direção ao seu centro. A Terra cria ao seu redor um campo gravitacional. Este campo atua sobre qualquer objeto, provocando o seu movimento de queda. Quando um corpo está sob ação exclusiva do campo gravitacional ele experimenta uma aceleração que não depende do tipo de material nem do estado físico do corpo. (pág. 72)
\end{quote}

\begin{figure*}[ht]
\centering
\includegraphics[scale=0.07]{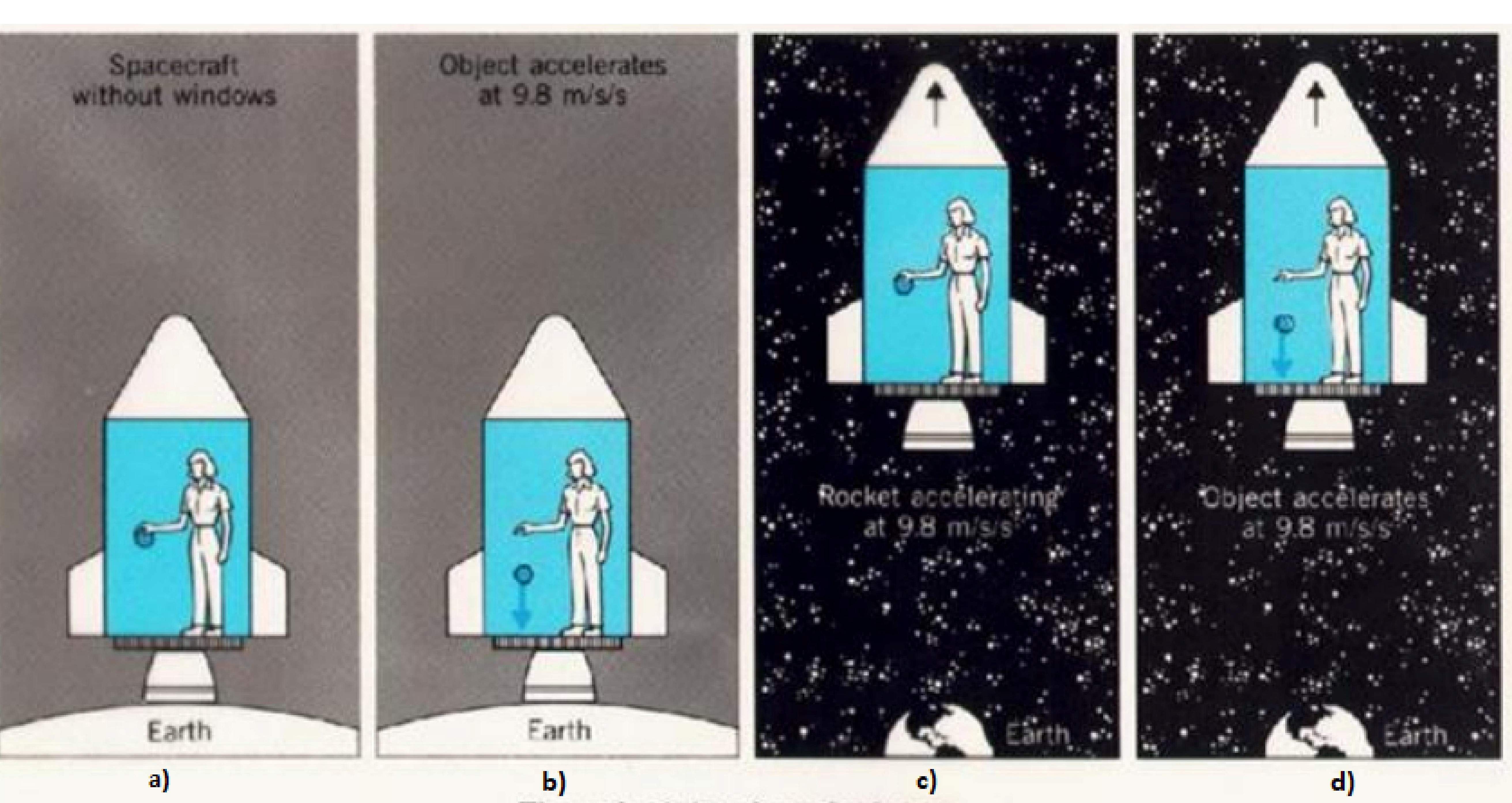}
\caption{A figura mostra um observador dentro de uma nave espacial sem janelas. Se o observador solta um objeto (quadros a e c), ele não irá saber a diferença entre está parado na superfície da Terra (quadro b), onde a aceleração gravitacional é 9,8 m/$s^2$, ou no espaço sendo acelerado (quadro d) para cima a 9,8 m/$s^2$. (Imagem modificada a partir da original disponível publicamente em http://www.if.ufrgs.br/~thaisa/wp-content/uploads/2017/03/Gravitação-e-princípio-da-equivalência.pdf)}
\label{equivalencia}
\end{figure*}

Isso parece um pouco não tão alarmante, mas quando pensamos na epistemologia da física, existe uma grande diferença entre o conceito de força e conceito de campo. A Relatividade Geral precisava ser compatível a essa nova ideia e explicar além da mecânica newtoniana. É nesse entendimento de interação de campo gravitacional sobre corpos que Einstein propôs o \textit{“Princípio da Equivalência”}, mostrando que corpos sobre efeitos gravitacionais em queda livre, ou seja, referenciais acelerados, não sentiam o seu peso, ou a influência direta da gravitação. Em outras palavras, Einstein fez a correlação ou a equivalência entre massa gravitacional e massa inercial mostrando que elas são proporcionais. O leitor pode elucidar melhor a ideia do princípio da equivalência observando a Fig.~\ref{equivalencia}.

Através da mecânica newtoniana, podemos relacionar a massa inercial, $M_{iner.}$, de um corpo com a sua aceleração, $a$, através da equação

\begin{eqnarray}
M_{iner.} = \frac{F}{a}\, ,
\label{massin}
\end{eqnarray}
onde $F$ é a resultante das forças que atuam no corpo.

Já para corpos em queda sob efeito gravitacional, sabemos que sofrem uma influência correlacionada com a gravidade como sendo o único ente acelerador, assim obtendo

\begin{eqnarray}
M_{grav.} = \frac{F}{g} \\
F = M_{grav.} \times g \, .
\label{force}
\end{eqnarray}

Substituindo a equação (\ref{force}) na equação (\ref{massin}), tem-se

\begin{eqnarray}
a = \frac{M_{grav.}}{M_{iner.}}\times g \, .
\end{eqnarray}

Considerando então o princípio da equivalência, e que num campo gravitacional a aceleração de um corpo independente da sua estrutura ou estado físico, é fácil compreender que $a=g$, e para isso ocorrer

\begin{eqnarray}
\frac{M_{grav.}}{M_{iner.}} = 1 \\
M_{grav.} = M_{iner.} \, .
\end{eqnarray}

Essa foi a porta de entrada da inclusão da gravitação na Relatividade.

Outro ponto que contribuiu para que a Relatividade Geral fosse formulada foi a tentativa de trazer uma covariância às equações que regiam o movimento dos corpos em uma única equação, ou um conjunto de equações que juntas explicariam tudo, mostrando que as leis da física são as mesmas em qualquer referencial e não mais apenas em referenciais inerciais como a Relatividade Especial trazia. Esse foi o avanço considerável entre a Relatividade Especial e a Geral, a covariância de ideias. É preciso também mencionar os efeitos de toda essa idealização; assim como a Relatividade Especial trouxe seus efeitos relativísticos com as transformações de Lorentz, a Relatividade Geral também trouxe fenômenos interessantes, sendo o mais intrigante e revolucionário, a curvatura do espaço-tempo.

\begin{figure}[!h]
\centering
\includegraphics[scale=0.043]{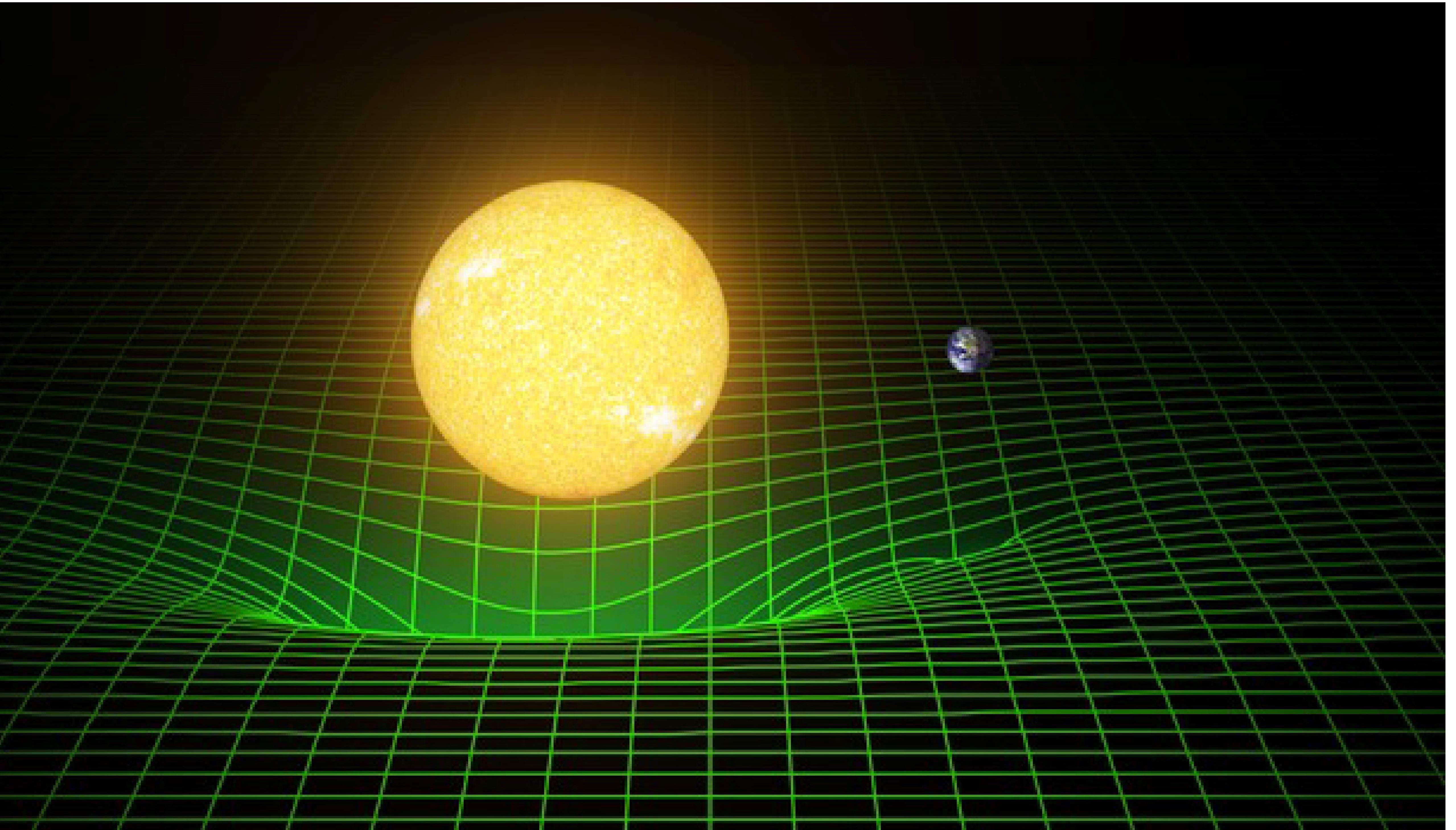}
\caption{Curvatura do espaço-tempo devido à massa do Sol e da Terra. (Imagem disponível publicamente em https://www.ligo.caltech.edu/WA/image/ligo20160211e)}
\label{curvatura}
\end{figure}

A curvatura do espaço-tempo (cf. Fig.~\ref{curvatura}) é outro efeito relativístico que mostra que corpos supermassivos através de relações gravitacionais curvam o espaço-tempo a sua volta, sendo esse, o primeiro ponto da fundamentação do objetivo principal deste trabalho que são as ondas gravitacionais. 

A ideia de curvatura do espaço trouxe outro choque na comunidade acadêmica da época. Saindo das concepções da geometria euclidiana, conhecida a mais de dois milênios, e entrando agora em uma nova geometria, dessa vez curvada, chamada de \textit{“Geometria não-Euclidiana”}, desenvolvida de forma independente pelos matemáticos Jamos Bolyay, Nicolai Lobachevsky, Carl Gauss e Bernhard Riemann~\cite{Riemann}, é então necessária para fazer jus a esse efeito relativístico da teoria. Com essa nova concepção geométrica do espaço, finalizava o ponto chave entre a teoria “einsteiniana” e a teoria newtoniana. Pois agora o movimento de planetas não está mais fundamentado no conceito de força gravitacional como ligação direta, mas sim pelo próprio movimento dos corpos que são pré-definidos devido a curvatura do espaço pela a matéria e energia. No caso do nosso sistema, devido ao o Sol, como enfatizado na Ref.~\cite{Peruzzo}, em dizer que “os planetas não giram em torno do Sol por causa de alguma força misteriosa, mas porque esta é a trajetória determinada pela geometria do espaço-tempo”, e também segundo o físico John Wheleer em uma carta sobre a Relatividade, menciona que "a matéria-energia diz ao espaço-tempo como se curvar, e o espaço-tempo diz a matéria-energia, como se mover".

Em 1916, Einstein formalizava matematicamente a curvatura do espaço-tempo através de um conjunto de equações de campo, conhecidas como \textit{"equações de Einstein"}, escritas num espaço quadri-dimensional de Riemann na forma~\footnote{Os índices de Lorentz que aparecem na equação possuem os valores $\mu, \nu, \ldots = 1,2,3,4$.}

\begin{eqnarray}
R_{\mu \nu} - \frac{1}{2}g_{\mu \nu}R = \kappa T_{\mu \nu} \, ,
\label{eqeinstein}
\end{eqnarray}
onde $\kappa$ é dado por

\begin{eqnarray}
\kappa = \frac{8\pi G}{c^4},
\end{eqnarray}
sendo $G = 6,672 \times 10^{-11} N \cdot m^2/kg^2$ a constante gravitacional e $c = 2,997 \times 10^8 m/s$ a velocidade da luz no vácuo. Segundo os autores da Ref.~\cite{Peruzzo}, “esta é uma das mais compactas e poderosas equações dentro da Física”.

Seguindo a descrição das Refs.~\cite{Cattani2,Maluf}, temos que do lado esquerdo da equação (\ref{eqeinstein}), o termo $R_{\mu \nu}$ representa o \textit{tensor de curvatura} de Ricci definido por

\begin{eqnarray}
R_{\mu \nu} = R_{\nu \mu} = \partial_{\sigma}\Gamma_{\mu \nu}^{\sigma} - \partial_{\nu}\Gamma_{\sigma \mu}^{\sigma} + \Gamma_{\sigma \tau}^{\sigma}\Gamma_{\nu \mu}^{\tau} - \Gamma_{\nu \tau}^{\sigma}\Gamma_{\sigma \mu}^{\tau} \, ,
\label{tensorricci}
\end{eqnarray}
sendo $\Gamma$ os chamados \textit{símbolos de Christoffel} dados por

\begin{eqnarray}
\Gamma_{\mu \nu}^{\sigma} =  \frac{1}{2}g^{\sigma \lambda} \left(\partial_{\mu}g_{\nu \lambda} + \partial_{\nu}g_{\mu \lambda} - \partial_{\lambda}g_{\mu \nu}  \right) \, .
\end{eqnarray}

O $g_{\mu \nu}$ é um dos termos mais conhecidos no que se refere a matemática relativística e representa o tensor métrico do espaço. Já o escalar $R$ é conhecido como escalar de curvatura ou invariante de curvatura do espaço e é definido como

\begin{eqnarray}
R = g^{\mu \nu} \left( \partial_{\sigma}\Gamma_{\mu \nu}^{\sigma} - \partial_{\nu}\Gamma_{\sigma \mu}^{\sigma} + \Gamma_{\sigma \tau}^{\sigma}\Gamma_{\nu \mu}^{\tau} - \Gamma_{\nu \tau}^{\sigma}\Gamma_{\sigma \mu}^{\tau}\right) \, .
\end{eqnarray}

Finalmente, $T_{\mu \nu}$ refere-se ao tensor energia-momento da matéria. Ele carrega toda a informação de massa, energia e momento de um sistema a ser estudado. 

Podemos observar que as equações de campo de Einstein formam um sistema de dez equações de derivadas parciais, não lineares e de segunda ordem, caracterizando a complexidade da matemática utilizada para a fundamentação da teoria da Relatividade Geral, com estruturas e conceitos um pouco mais rebuscados, que até o próprio Einstein se refere a essa “matematização” da sua teoria dizendo que depois que os matemáticos entraram em sua teoria, nem ele a entenderia mais.

Devemos enfatizar que o motivo da apresentação da matemática introdutória da Relatividade Geral neste trabalho não é de caráter descritivo e minucioso, mas apenas em caráter de conhecimento e familiarização da teoria de campo, necessária ao leitor para compreender que as equações básicas que preveem as ondas gravitacionais podem ser obtidas a partir das equações de campo de Einstein. Um aprofundamento na discussão matemática acerca das ondas gravitacionais pode ser encontrado nas Refs.~\cite{Cattani, Cattani2, Maluf, Cattani3, Cattani4}.

\subsection{Contexto histórico e o desenvolvimento das OG}

O conceito de ondas gravitacionais foi proposto em 1916, onde Albert Einstein publicara o seu primeiro artigo sobre a ideia da sua existência como uma das consequências direta do entendimento sobre a teoria da Relatividade Geral. O problema no que se propõe a seguinte afirmativa, \textit{“ondas gravitacionais propostas em 1916”}, é dar a ideia de que as ondas gravitacionais já estavam de fato fundamentadas e prontas para enfrentar as discussões na comunidade científica, no entando, a afirmação não é correta.  Assim como a ideia da relatividade, as ondas gravitacionais não eram um conceito tão novo quanto se parece para a época, já sendo discutido desde o século 18 mas encontrando fundamentos teóricos pra sua existência apenas em meados do século 20, surgindo como um resultado inerente ao desenvolvimento da Relatividade Geral de Einstein. Com isso, o conceito das OG surgiu como resposta para algumas questões que a teoria gravitacional newtoniana não era capaz de prever e explicar como, por exemplo, o prolongamento do periélio de Mercúrio.
  
O problema de Mercúrio sempre foi uma “pedra no calçado” de vários físicos que adentravam aos estudos gravitacionais. Por ser um sistema de dois corpos e Mercúrio o planeta mais próximo do Sol com orbitas de maior excentricidade, exatamente 0,2056, as relações gravitacionais envolvidas nesse sistema são mais atenuantes do que em qualquer outro sistema de dois corpos, daí então a sua complexidade e visibilidade por parte de estudos. Segundo Kennefick~\cite{Kennefick}, em 1908 o físico Henry Poincaré já mencionava a existência de ondas gravitacionais, sob a influência da teoria eletromagnética de Maxwell, como resposta para a discrepância no prolongamento do periélio de Mercúrio em forma de perda energética, afirmando que “(...)  a emissão de ondas gravitacionais a partir da órbita deste planeta interno que se deslocava rapidamente estava removendo energia suficiente de seu movimento para aparecer na forma do deslocamento do periélio~\footnote{Versão livre do original, \textit{"That the emission of gravitational waves from the orbit of this quickly moving inner planet was removing sufficient energy from its motion as to show up in the form of the perihelion shift".} (pag. 38)}”, fazendo ainda mais uma analogia com cargas elétricas aceleradas que liberam radiação.

Vale mencionar que a utilização do termo de ondas gravitacionais como resposta ao avanço do periélio de Mercúrio foi debatido anos depois e desmistificado por Einstein em seus trabalhos de escolha de coordenadas para o estudo de ondas gravitacionais.

As ondas gravitacionais, também chamadas de radiação de amortecimento ou ondas de aceleração, surgiram inicialmente em dois âmbitos: 

\begin{enumerate}
\item Com o propósito de explicar as adversidades nos estudos de Mercúrio e seu periélio, como também explicar qualquer influência gravitacional que se propague pelo espaço, já que a ideia de uma força gravitacional instantânea de Newton estava perdendo sua concepção, dando a ideia de que se ondas gravitacionais se propagam, deveriam assim fazer com velocidade finita, pois nada poderia superar a velocidade da luz;

\item Conseguir fazer uma correlação entre gravidade e eletromagnetismo. A eletricidade e o magnetismo já haviam sido organizadas anos atrás por James Clerk Maxwell, gerando então, o Eletromagnetismo. Um passo imensurável no entendimento das forças que regem o Universo. Einstein também compartilhava dessa ideia de que era possível mensurar todo o nosso conhecimento físico em uma única força que pudesse explicar tudo. Morreu tentando fazer isso; e as ondas gravitacionais, como já mencionado, nada mais são do que uma tentativa de fazer essa união entre eletromagnetismo e gravitação.
\end{enumerate}

Com Maxwell, não veio apenas o Eletromagnetismo ou a quantização da velocidade da luz, mas também adentrou ao meio físico o ideal de campo. Campo como entidade física, mesmo com sua complexidade no quesito entendimento filosófico, pode ser bem aplicado em diversas situações físicas, dando novas explicações de funcionamento. Esse mesmo ideal, veio parar na gravitação, saindo do termo força e entrando no termo campo, e aí onde está a primeira base analógica entre eletromagnetismo e gravitação que já era discutido no ano de 1908~\cite{Kennefick}.

Campos eletromagnéticos podem gerar ondas eletromagnéticas, não poderia então campos gravitacionais gerar ondas gravitacionais? Essa é a pergunta primordial da sua fundamentação e esse também é o primeiro comparativo, a primeira analogia para então fundamentar as ondas gravitacionais e dar uma fixada na Relatividade Geral de Einstein. Durante os anos de formalização das ondas gravitacionais, muitas analogias foram fundamentais para servirem de base consistente no entendimento e visualização da existência de ondas gravitacionais, isto será discutindo com mais detalhes na próxima seção.

Com o avanço do entendimento das ondas gravitacionais, passando por descrenças e crenças, dois problemas aparecem como decisivos para a estabilização da Teoria: 

\begin{enumerate}
\item Como essas ondas se propagam?, visando não somente a ideia matemática (como equações de ondas), mas uma forma conceitual;

\item E qual a fonte dessa irradiação gravitacional? Como ela funciona ou deveria funcionar?
\end{enumerate}

O segundo problema apresentado foi um dos mais complexos a ser examinado, pela dificuldade observacional e experimental das ondas gravitacionais na época, o que levou muitas vezes as descrenças e descartes dessa teoria. Esses são os dois pontos cruciais que levaram uma gama de físicos, matemáticos e engenheiros a buscarem entender um pouco mais sobre as resoluções matemáticas de ondas aplicadas às ondas gravitacionais e a entender sobre sistemas binários, como estrelas e buracos negros, fortes concorrentes a serem capazes de formular ondas gravitacionais, focando um pouco na solução de Schwarzschild para as equações de Einstein.

As ondas gravitacionais só foram começar a ser de fato fundamentadas de forma fixa e predominante em 1937, 21 anos depois da sua apresentação, com uma contribuição importante de Robert Oppenheimer que encontrou discrepâncias nas soluções de Leopold Infeld, assistente de Einstein em meados dos anos 1930, trazendo uma solução final para ondas planas e infinitas, necessárias para fazer das OG um fato existente e consequentemente um progresso inigualável na teoria. Fora Oppenheimer, outros físicos importantes se envolveram na construção filosófica, matemática e física do ideal gravitacional, alguns de forma a negar e comprovar a sua não existência, como Max Abraham, considerado como o primeiro cético à teoria, por publicar um trabalho referindo-se às OG que não poderia desempenhar nenhum papel significativo na teoria relativística da gravitação~\cite{Kennefick} e outros a acreditar na teoria da Relatividade e consequentemente nas suas implicações.

O número de opositores à teoria da Relatividade sempre sucumbiu aqueles que a apoiavam, mesmo após anos de estudos e desenvolvimento da teoria, muitos daqueles que eram opositores acabaram por se tonar fontes indispensáveis no desenvolvimento das OG. Alguns dos exemplos são os físicos Karl Schwarzschild, Gunnar Nordström, Leopold Infeld, Wolfgang Pauli, Max Born e vários outros. Um dos grandes opositores da ideia da existência das OG foi o astrofísico Arthur Stanley Eddington, que questionava uma série de problemas que de fato existiam nos primórdios da idealização de Einstein, como a propagação de onda e energia, a “incapacidade” de demonstração experimental para a época e principalmente no que diz respeito a velocidade de propagação da onda proposta por Einstein. Eddington foi tão grande opositor, da ideia de onda gravitacional se propagando mediante corpos massivos no espaço-tempo, que formulou umas das frases mais icônicas e irônicas na época da formulação das OG, em mencionar que as ondas gravitacionais se propagavam na “velocidade do pensamento”. Esse questionamento é mostrado em um artigo publicado pelo próprio Eddington~\cite{Eddington} em 1922.

Essa discussão fervorosa, perdurou por um longo tempo, dando a Eddington, segundo o autor da Ref.~\cite{Kennefick}, o patamar de um dos primeiros céticos fervorosos da teoria, gerando uma verdadeira "perseguição" a Einstein. Eddington não estava errado em questionar o desenvolvimento das OG, pelo contrário, se tornava um impulso para que inúmeros problemas com relação a fundamentação da teoria fossem superados um a um, como menciona Kennefick~\cite{Kennefick}. Como já mencionado no decorrer de todo processo construtivo das OG, a maior dificuldade se deu na sua comprovação experimental, processo crucial na validação teórica, onde a menção inicial seria de que o modelo proposto para supostamente irradiar as ondas mencionadas seriam um sistema binários de estrelas, mas depois sendo mencionado a possibilidade de buracos negros também irradiarem, como proposto por Schwarzschild. Entendia-se que por estarem tão distantes em escala astronômica, as ondas propagadas pelo espaço que chegariam até a Terra, seriam de dimensões tão desprezíveis quanto um alfinete em um palheiro, como apresenta Saa~\cite{Saa} reafirmando que “as estimativas iniciais, feitas pelo próprio Einstein, apontavam corretamente que as amplitudes típicas das ondas gravitacionais seriam minúsculas, talvez não detectáveis na prática”; e outros problemas como: métrica de espaço-tempo, tipo de coordenadas fundamentadas, propagação da onda, funcionamento da fonte (sistema binário), que juntos fizeram até do próprio Einstein, cético da sua ideia, chegando a ponto de mencionar inúmeras vezes que as ondas gravitacionais não existiam como mostra uma carta~\cite{Born} em meados de 1936 escrita pelo próprio Einstein para Max Born, dizendo que “(...) cheguei ao interessante resultado de que as ondas gravitacionais não existem, embora tivessem sido assumidas como uma certeza para a primeira aproximação”~\footnote{Versão livre do original, \textit{"I arrived at the interesting result that gravitational waves do not exist, though they had been assumed a certainty to the first approximation".} (pag. 125)}.

Com a métrica utilizada por Einstein e depois de uma reviravolta com o uso de pseudo-tensores e resoluções de equações linearizadas espaço-temporais, três tipos de ondas são supostamente encontradas e que estão inerentes às ondas gravitacionais, as ondas Longitudinais Longitudinais (LL), Longitudinais Transversais (LT) e as Transversais Transversais (TT) que serão mais detalhadas nas próximas seções deste trabalho. A problemática com o tipo de coordenada utilizada para formalizar as ondas gravitacionais tem uma relação direta com os 3 tipos de equações de onda encontradas por Einstein, pois a sua mudança foi o fator responsável por trazer essa variância dos 3 tipos de ondas. Vale a pena frisar um pouco mais sobre a disputa de coordenada usada por Einstein para viabilizar as OG, pois foi um fator importantíssimo para o andamento da teoria e consequentemente sua fortificação~\cite{Kennefick}.

Inicialmente, Einstein apresenta as ondas gravitacionais trazendo consigo, pela primeira vez, as aproximações linearizadas das equações de campo advinda de teorias anteriores, sendo que a escolha de coordenadas iniciais, as coordenadas unimodulares, deixava os cálculos mais simplórios e fáceis de manipular com um fator de métrica definido por $\sqrt{-g}=1$. Na tentativa de usar as mesmas coordenadas adotadas na aproximação pós-newtoniana com uma falsa impressão de que os cálculos presentes seriam simplificados para o estudo das OG, e na esperança de fazer menção da sua teoria gravitacional ser compatível com a newtoniana em algum aspecto, Einstein comete um erro; com o apoio de Schwarzschild, observa uma discrepância para esse tipo de problema gravitacional, onde a coordenada adotada gerava parâmetros matemáticos inconsistentes, sendo essa umas das primeiras provas concretas de quebra da analogia das ondas eletromagnéticas e a gravitação. Em 1916 Einstein comenta esse fator a Schwarzschild através de carta. Meses após a sua convicção de que ondas gravitacionais não poderiam existir por inconsistência matemáticas, ele mudou a sua forma de analisar as propriedades do estudo, utilizando dessa vez não de uma aproximação pós-newtoniana, mas se utilizando de concepções adotadas na Relatividade Especial e sua aproximação linearizada, também chamada de coordenadas isotrópicas, introduzindo o ideal de OG na Relatividade Geral, sobre influência do astrônomo Willem de Sitter, grande influenciador do desenvolvimento das ondas gravitacionais.

A explicação da utilização de fatores linearizados em uma teoria cujo principal componente é a não linearidade das equações, como a Relatividade Geral, se dá em uma abordagem um pouco diferente sobre as OG na teoria geral. Sendo a teoria da Relatividade Geral embasada em uma geometria não euclidiana, ou seja, espaço-tempo curvos devido a corpos massivos, e a radiação gravitacional como de uma amplitude mínima, quase desprezíveis, pode-se adotar, para esse caso, o espaço não como curvo, mas sim planificado já que as curvaturas causadas pelas ondas gravitacionais seriam mínimas, dando então uma boa chance de encontrar equações condizentes pela sua simplicidade. Com a mudança de aproximações da pós-newtoniana para a Relatividade Especial, em 22 de junho de 1916 em uma carta amigável a Sitter, Einstein abandona de vez a coordenada unimodular de fator $\sqrt{-g}=1$, devido a utilização de potenciais retardados trazendo uma nova aproximação a teoria eletromagnética de campo~\cite{Kennefick}.

Em Janeiro de 1918, Einstein encontra e conserta um erro de derivação em seus pseudos-tensores utilizados para o cálculo das energias em suas equações publicadas em um artigo em 1916, gerando então novas formulações quadripolares para as ondas gravitacionais, sendo explicada como a “fórmula de radiação revisada que expressa a quantidade de energia irradiada por uma fonte de ondas gravitacionais”~\footnote{Versão livre do original, \textit{"revised formula that expresses the amount of energy radiated by a source of gravitational waves".} (pag. 71)}~\cite{Kennefick}, com ajuda do físico irlandês Gunnar Nordström que segundo Kennefick~\cite{Kennefick} era considerado como “um dos primeiros pioneiros da teoria da Relatividade, que foi, de fato, o primeiro a desenvolver uma teoria relativista totalmente auto consistente da gravidade, um ano antes da Einstein”~\footnote{Versão livre do original, \textit{"one of the earlier pioneers of relativity theory, who was in fact the first to develop a fully self-consistent relativistic theory of gravity, a year before Einstein’s".} (pag. 71)}. Surge de forma incisiva a ideia de formulações quadripolares, que são importantes para o entendimento de fonte das OG. 

Com o avanço da teoria e a “crença” de Einstein em uma teoria unificadora, o mesmo leva o estudo das ondas gravitacionais em meados dos anos 1950, a uma tentativa de unificação entre a estrutura quântica e a gravitação, com relação ao seu pensamento e estudo sobre elétrons atômicos, formalizando o ideal, nomeado anos depois de gravitação quântica, sendo também responsável por trabalhos e estruturação das OG.
	
Mesmo com todo esforço empreendido, avanços e perdas, as OG só foram ter uma estrutura fixada e predominante nas décadas de 1960 e 1970, onde a sociedade tem como aliado o desenvolvimento tecnológico e avanços da engenharia de detectores, necessários para trazer um avanço inigualável tanto na formalização teórica, quanto e principalmente na realização experimental, fazendo dessa teoria algo real e imponente na sociedade científica, abrindo novas portas para o conhecimento.

\subsection{Processos analógicos para o entendimento das OG}

Nesta seção decorreremos sobre a analogia da formalização científica das OG, enfatizando a possibilidade delas serem a porta de entrada para a unificação do eletromagnetismo e gravitação, sendo amplamente presente e fundamental no desenvolvimento da teoria das OG. O processo de analogia empregada em teorias científicas sempre foi um dos mecanismos mais utilizados na caracterização e entendimento da mesma, servindo como terreno fértil tanto para a descoberta quanto para a controvérsia, tendo como objetivos primordiais:

\begin{enumerate}
\item Trazer um embasamento inicial em conceitos já estruturados e definir modelos teóricos viáveis, contribuindo assim para o desenvolvimento da sua formalização;

\item Gerar uma futura unificação de conceitos físicos na tentativa de encontrar padrões para o desenvolvimento da “teoria de tudo”.
\end{enumerate}

São incontáveis os processos científicos embasados por analogias iniciais como parâmetro moldador a fim de chegar a uma teoria concreta e fundamentada, como foi o caso da formalização da teoria cinética dos gases, em 1738, e anos mais tarde a tentativa de modelagem do éter luminífero, no séc. XIX. 

A analogia não pode ser entendida como uma menção igualitária e simétrica, mas sim como uma comparação entre dois sistemas de ideias. Até hoje existem muitos embates sobre o significado do termo “analogia” e o que de fato ela pode proporcionar ao entendimento de entes físicos, mas é incontestável a sua eficácia na idealização de parâmetros para o estudo do desconhecido e aprimoramento do conhecido. 

Segundo Duarte~\cite{Duarte}, “a analogia não pressupõe, portanto, a existência de uma igualdade simétrica, mas antes uma relação que é assimilada a outra relação, com a finalidade de esclarecer, estruturar e avaliar o desconhecido a partir do que se conhece”. 

Para enfatizarmos ainda mais esse processo, podemos mencionar Thomas Kuhn, um dos mais respeitados filosóficos científicos do séc. XX, é um grande defensor da ideia de analogias presentes na ciência como uma forma de elucidação de ideias e formulação de teorias primitivas, sendo a partir de então mais palpáveis e lapidáveis. 

A presença da analogia no campo de estudo das ondas vem percorrendo um caminho longo desde o século IV a.C. com fundamentos de Aristóteles e demais filósofos gregos e romanos sobre o som, que segundo os mesmos, era uma onda que viajava pelo ar, em analogia com ondas formadas na superfície da água em um copo. Passando para o desenvolvimento físico do século XVIII, com trabalhos de Isaac Newton e Chistiaan Huygens, e século XIX com a proposição da luz como uma onda devido a analogia com a propagação do som, chega-se por fim ao século XX com os teóricos relativistas levantando a ideia analógica entre o eletromagnetismo e a gravitação sugerido a presença de ondas em uma teoria da gravitação. 

Esse foi o primeiro momento da presença de analogias para embasar a ideia das ondas gravitacionais, levantando indagações que, se o eletromagnetismo tem uma influência de propagação das ondas eletromagnéticas pode então ter a gravitação a influência de ondas gravitacionais? Em caso particular do desenvolvimento das OG a utilização de ideias analógicas teve um caráter maior, não de explicar e entender a teoria, mas sim de correlacionar uma matemática e de prevê-la com base nos avanços da relatividade, se tornando ainda mais necessária devido à falta de comprovação experimental, sendo reafirmada por Kennefick~\cite{Kennefick} em dizer que “começando com sua abordagem, os teóricos da relatividade analisaram várias analogias com o campo eletromagnético enquanto tentavam construir uma teoria das ondas gravitacionais na ausência de evidências experimentais”~\footnote{Versão livre do original, \textit{"Beginning with his approach relativity theorists looked to various analogies with the electromagnetic field as they attempted to construct a theory of gravitational waves in the absence of experimental evidence".} (pag. 71)}, que somente vai ser dada anos depois, principalmente quando se refere a analogia imediata empregada a teoria gravitacional com o eletromagnetismo de Maxwell, já mencionado em seções anteriores. Segundo Infeld~\cite{Infeld}, assistente de Einstein, a existência das ondas gravitacionais pode ser deduzida da Relatividade Geral, e as ondas eletromagnéticas deduzidas da teoria de Maxwell.

Outro ponto importante para a firmação da analogia com o eletromagnetismo se deu pela ideia da velocidade de propagação de uma onda gravitacional que deveria então respeitar intrinsecamente o dito pela teoria eletromagnética em que nada poderia se propagar com velocidade maior do que a luz. Questão debatida mais tarde por Eddington. Vale ressaltar que a ideia da gravidade se propagar como ondas e com velocidades finitas foi o ponto crucial para a reformulação da teoria gravitacional de Newton. O choque com a ideia eletromagnética que até então parecia tão promissora para uma dada unificação, muda quando começa a se tratar da ideia de como deveria funcionar a fonte emissora dessas ondas gravitacionais, gerando ideais monopolares, dipolares e quadripolares. 
	
\begin{figure}[!h]
\centering
\includegraphics[scale=0.30]{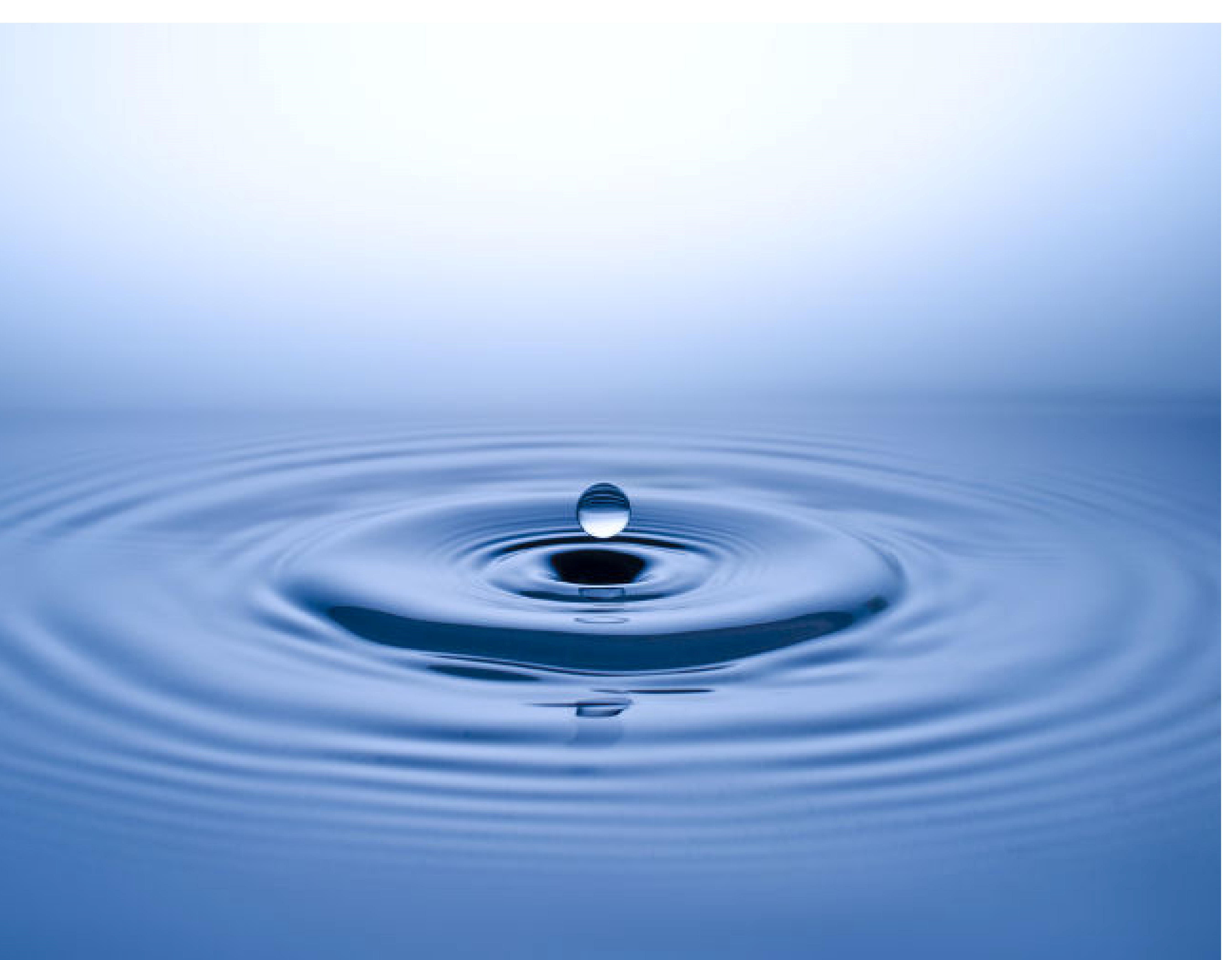}
\caption{Surgimento de ondas devido a interação de uma gota com a superfície da água. (Imagem disponível publicamente em https://brasilescola.uol.com.br/fisica/a-classificacao-das-ondas.htm)}
\label{ondasagua}
\end{figure}

\begin{figure}[!h]
\centering
\includegraphics[scale=0.025]{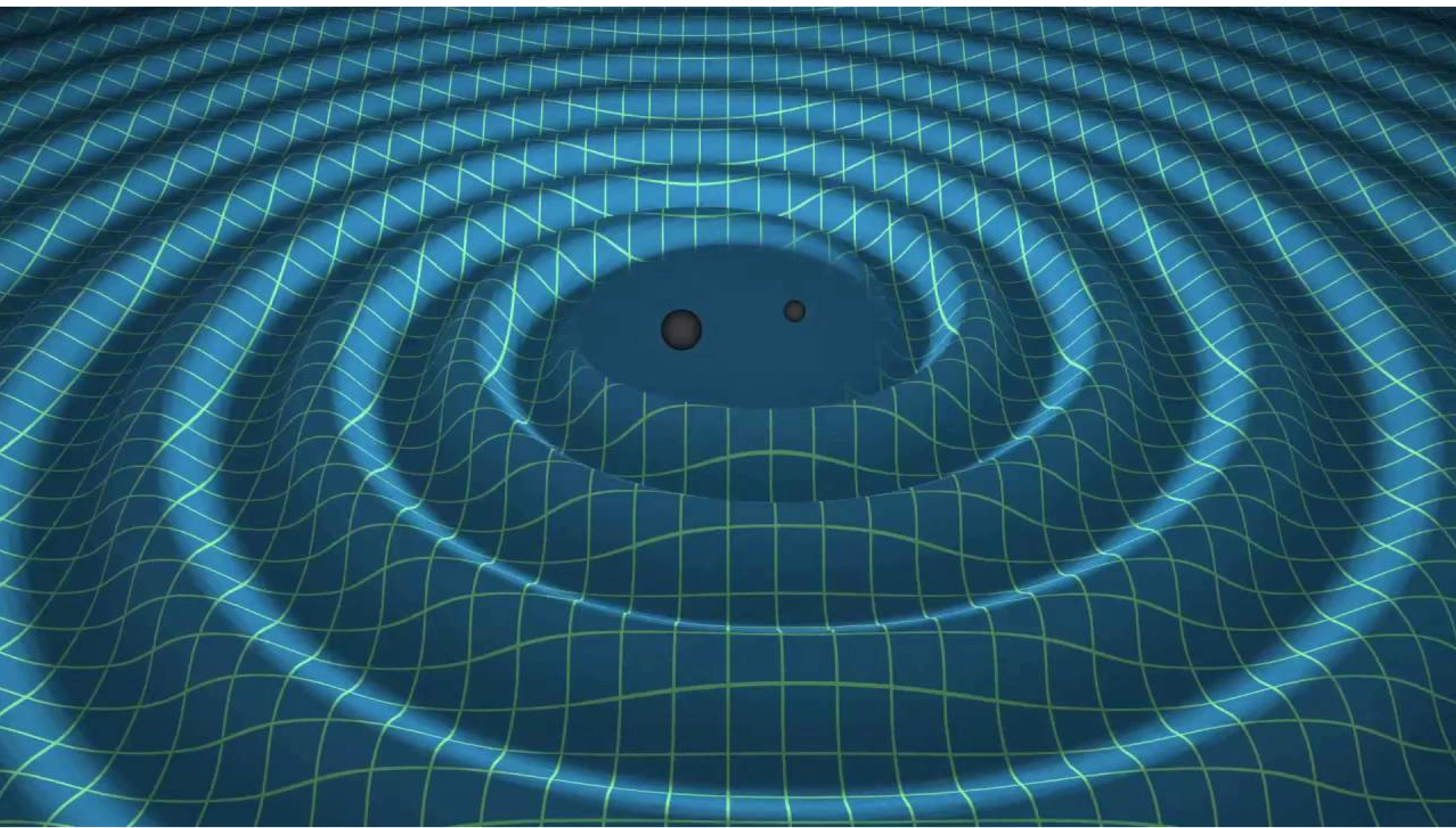}
\caption{Representação das ondas gravitacionais devido à fusão de dois buracos negros. (Frame retirado do video de simulação em https://www.ligo.caltech.edu/mit/video/ligo20160615v1)}
\label{og}
\end{figure}

Em meados do século XIX a menção “ondas gravitacionais” era incrivelmente relacionada ao estudo das oscilações em superfícies da água, dada em razão do cálculo do próprio peso da água, sendo mais tarde o termo utilizado para se atribuir a ideais gravitacionais~\cite{Kennefick}. Abre-se então uma outra ideia analógica com base nas perturbações na superfície da água devido a propagação de ondas, como podemos observar na Fig.~\ref{ondasagua}, usada como uma demonstração mais simplória e didática sobre o funcionamento de OG. Com o desenvolvimento da teoria, essa analogia nos leva ao modelo mais atual da propagação das ondas gravitacionais como podemos observar na Fig.~\ref{og}.

Essa analogia foi uma das mais fundamentais e importantes para o desenvolvimento da teoria, pois vai ser motivo de discussão sobre a utilização de coordenadas isotrópicas por Einstein, já relatadas anteriormente, e principalmente do funcionamento da fonte emissora. A analogia com as perturbações na água tomou outro enfoque quanto à discussão sobre o espaço de propagação das OG, quando Einstein entende a dificuldade da observação experimental das suas ondas, justamente por possuírem uma amplitude muito baixa para serem detectadas, gerando equações de ondas planas, por conterem baixas amplitudes, as discrepâncias e perturbações no espaço-tempo seriam tão mínimas que poderiam ser consideradas como planas. Essa nova contradição, também já relatado em seções anteriores, tende a ser analisada segundo a analogia das ondas na água.

\begin{figure}[!h]
\centering
\includegraphics[scale=0.70]{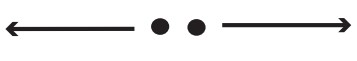}
\caption{Representação de uma interação dipolar. (Figura retirada da Ref.~\cite{Kennefick})}
\label{dipolo}
\end{figure}

Essa associação foi de suma importância, pois levou à discussão do funcionamento da fonte emissora das ondas gravitacionais, ponto crucial debatido exaustivamente por Einstein para a fundamentação final da teoria, sendo que ao mesmo tempo que discute algo tão primordial para o desenvolvimento das OG. Aponta também um dos primeiros desequilíbrios entre a analogia eletromagnética e gravitacional, para o entusiasmo dos céticos que não aceitavam tal analogia como Arthur Eddington, Nathan Rosen e principalmente Hermann Bondi por questionar a liberação de energia pelo sistema dipolar e quadripolar. Já era de conhecimento que as radiações eletromagnéticas trabalhavam com sistemas dipolares (cf. Fig.~\ref{dipolo}), onde as perturbações são geradas e se movem ao longo de uma linha, possuindo variação em apenas um eixo, não possuindo uma simetria esférica como os monopolos, mas que ainda possui uma simetria axial, gerando a chamada radiação dipolar. 

\begin{figure}[!h]
\centering
\includegraphics[scale=0.60]{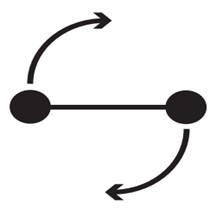}
\caption{Representação de uma interação quadripolar. (Figura retirada da Ref.~\cite{Kennefick})}
\label{quadrupolo}
\end{figure}

Um exemplo didático se dá no funcionamento de antenas de rádio que se relacionam por um dado dipolo. Esse tipo de sistema foi em primeira instancia recusado por um apanhado de físicos quando aplicado a sistemas gravitacionais, tanto por ser inaceitável a ideia de uma gravitação repulsiva, já que os corpos oscilariam em apenas um eixo e também por ferir a ideia de conservação, que no caso gravitacional, deveria se dar pela conservação do momento, onde cada massa acelerada deveria influir o mesmo momento, o que acabaria anulando qualquer influência de campo, não liberando, portanto, radiação. Isso gerou uma discussão fervorosa principalmente entre o maior cético da analogia Eletromagnetismo-Gravitação, Hermann Bondi e o maior defensor e pioneiro na detecção de OG, John Weeler e também com demais cientista. Uma discussão que teve seu início em 1914, devido a sistemas de estrelas binárias e se estendeu até 1970 aplicado a sistemas gravitacionais. Entendeu-se então que para sistemas gravitacionais onde massas estariam acelerando e que então deveria liberar energia, ponto em que Bondi foi rígido em questionar, só ocorreria em sistemas quadripolares (cf. Fig.~\ref{quadrupolo}) que se apresenta em um sistema com movimentos em torno de dois eixos separados, logo quatro polos, em simetria, muito utilizados na idealização de sistemas binários, a muito já estudados. A quantidade de polos presentes na geração da onda também explica o fato do porquê de OG terem amplitudes tão baixas "inviabilizando" a sua detecção, mostrando que de fato não poderia se apresentar em sistemas dipolares. A apresentação de sistemas quadripolares para OG deu abertura para Einstein formular suas equações de perda de energia em função do tempo com relação a fonte de emissão.

Mesmo tendo uma de suas primeiras contradições, a analogia Eletromagnetismo-Gravitação sempre foi defendida e apresentou uma série de regularidades que contribuíram para o desenvolvimento das OG, não sendo pelo fato de ter encontrado uma discordância que tal analogia perdeu sua credibilidade~\cite{Kennefick}.

\subsection{Ondas gravitacionais e a detecção}

Segundo o LIGO~\cite{ligo}, “as ondas gravitacionais são "ondinhas" no tecido do espaço-tempo causado por alguns dos processos mais violentos e energéticos do Universo”, em outras palavras, ondas gravitacionais são oscilações espaço-temporais ocasionadas por corpos super-massivos que ao se colidirem por perda energética, liberam energia em forma de ondas que curvam o espaço-tempo e se propagam na velocidade da luz. Essa é a forma mais simplória de se mencionar OG para o entendimento mútuo. 

Mesmo com todo avanço em cima da teoria ondulatória da gravitação, o maior problema da passagem das OG de "simples" teoria para se tornar algo aplicável e real, foi sempre a sua experimentação. Como já mencionado, essa era a “pedra” que Einstein sempre teve que carregar e que o levou muitas vezes a desacreditar na sua teoria, assim como também, era o argumento que muitos dos céticos utilizavam para desestimular aqueles que tanto se esforçavam para dar vida a crença nas OG. Era compreensivo a falta de preparo tecnológico para a detecção das OG na época, sabendo que as revoluções tecnológicas somente tomaram força apenas na metade do séc. XX, mas mesmo assim se tornou um grande empecilho na fundamentação da teoria.

Essa problemática começa a ter outro rumo a partir da década de 1960 e 1970, com Joseph Weber, físico americano e pioneiro na tentativa de detecção das OG, e seus alunos. Vale lembrar que nessa época as OG já possuíam uma estrutura matemática e física bem construída, logo faltava apenas a sua detecção. Com a iniciativa de Weber, começa-se uma verdadeira corrida tecnológica e científica na construção de dispositivos capazes de detectar as tão esperadas OG. Um período que foi essencial para o desenvolvimento tecnológico na Física. 

Outro nome icônico na busca pela detecção das OG de Einstein, é Kip Thorne, que segundo Kennefick, foi um dos pioneiros e um dos maiores impulsionadores na criação e manutenção do LIGO, responsável atualmente pela detecção direta das OG. O autor da Ref.~\cite{Kennefick} menciona algo interessante sobre a trajetória de credibilidade de Thorne, mencionando que em 1981, Thorne entra em uma aposta com um dos maiores críticos experimentais das OG, o astrônomo Jeremiah Ostriker, em que as OG seriam detectadas no final do séc. XX, o que de fato não ocorreu e Thorne perdeu a aposta.

Não desacreditando em seu trabalho, em 1992, o LIGO é implementado como um centro de pesquisa e detecção por Kip Thorne, Ronald Drever, físico escocês do CALTECH (California Institute of Technology), e Rainer Weiss, físico americano do MIT (Massachusets Institute of Technology). Juntos e com o apoio da NFS (National Science Foundation) e de outros laboratórios como o VIRGO Interferometer, constroem um dos mais emblemáticos e promissores dispositivos experimentais de meio bilhão de dólares, com o objetivo de detectar as famosas ondas gravitacionais. Após 100 anos da predição de Einstein da existência das OG e 24 anos de pesquisa e trabalho persistente, no dia 11 de fevereiro de 2016, o LIGO divulga para a mídia mundial a observação direta das OG em um artigo intitulado “Observation of Gravitational Waves from a Binary Black Hole Merger” publicado na \textit{Physical Review Letters} \cite{Abbott} entusiasmando a comunidade científica e trazendo luz para novos campos de estudos como a da Astronomia Gravitacional. A divulgação das detecções ocorreu em 2016, mas as detecções em si ocorreram em duas épocas, a primeira em 14 de setembro de 2015 e a segunda em 26 de dezembro de 2015, trazendo uma segurança a mais na confirmação das ondas gravitacionais.

\begin{figure}[!h]
\centering
\includegraphics[scale=0.60]{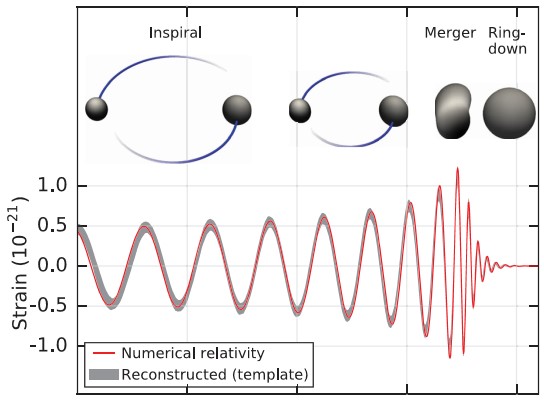}
\caption{Representação dos estágios de um BBN captado pelo LIGO com a respectiva estimativa da amplitude de deformação das OG. (Figura retirada da Ref.~\cite{Abbott})}
\label{bbns}
\end{figure}

A detecção que o LIGO conseguiu captar veio de um processo já esperado entre buracos negros há 1,3 milhões de anos-luz da Terra. Tal momento na história do desenvolvimento científico, ofereceu não só a comunidade científica provas da existência das OG como também bases reais da ação e existência de buracos negros. Já era predito que os únicos entes capazes de gerar OG que possam se propagar pelo espaço eram estrelas binárias ou um sistema binário de buracos negros, devido a sua elevada massa, como menciona os autores da Ref.~\cite{Cattani} e já citado na introdução deste trabalho. No caso da detecção, o fenômeno ocorreu entre buracos negros de massas respectivamente iguais a 36 e 29 vezes a massa do Sol passando por 3 estágios: \textit{inspiral}, \textit{merger} (ou \textit{plunge}) e por fim o \textit{ringdown} como esquematizado na Fig.~\ref{bbns}.

Com base na Ref.~\cite{Cattani}, podemos obter uma explicação, mais detalhada, sobre cada estágio do sistema binário apresentado na Fig.~\ref{bbns}. Temos que a fase \textit{Inspiral} é a fase inicial da vida de um sistema binário de buracos negros que oscilam um em relação ao outro, com baixa velocidade e com baixa produção de OG. O tempo de diminuição das órbitas é muito grande devido a pequena força de atração. Com o passar do tempo as órbitas vão se encolhendo, devido a perda energética e consequentemente aumenta a velocidade orbital e a liberação de OG fica mais evidente e ativa. A medida em que se aproximam a amplitude da onda atinge o seu máximo, esse é o segundo estágio, \textit{Merger}, que chega a atingir velocidades extremamente elevadas até o ponto que a perda energética sucumbe o sistema e acabam por se fundir em um só buraco negro atingindo uma espécie de estabilidade, essa é a terceira fase conhecida como \textit{ringdown}. Quando o \textit{ringdown} ocorre, uma gama de energia em forma de OG é liberada de forma violenta. No caso do sistema binário detectado pelo LIGO, o momento de \textit{ringdown} ocorreu em um buraco negro resultante de 62 vezes a massa do Sol, com uma liberação energética em forma de ondas, equivalente a 3 vezes a massa do Sol. 

\begin{figure*}[ht]
\begin{minipage}[b]{0.45\linewidth} 
\includegraphics[width=\linewidth]{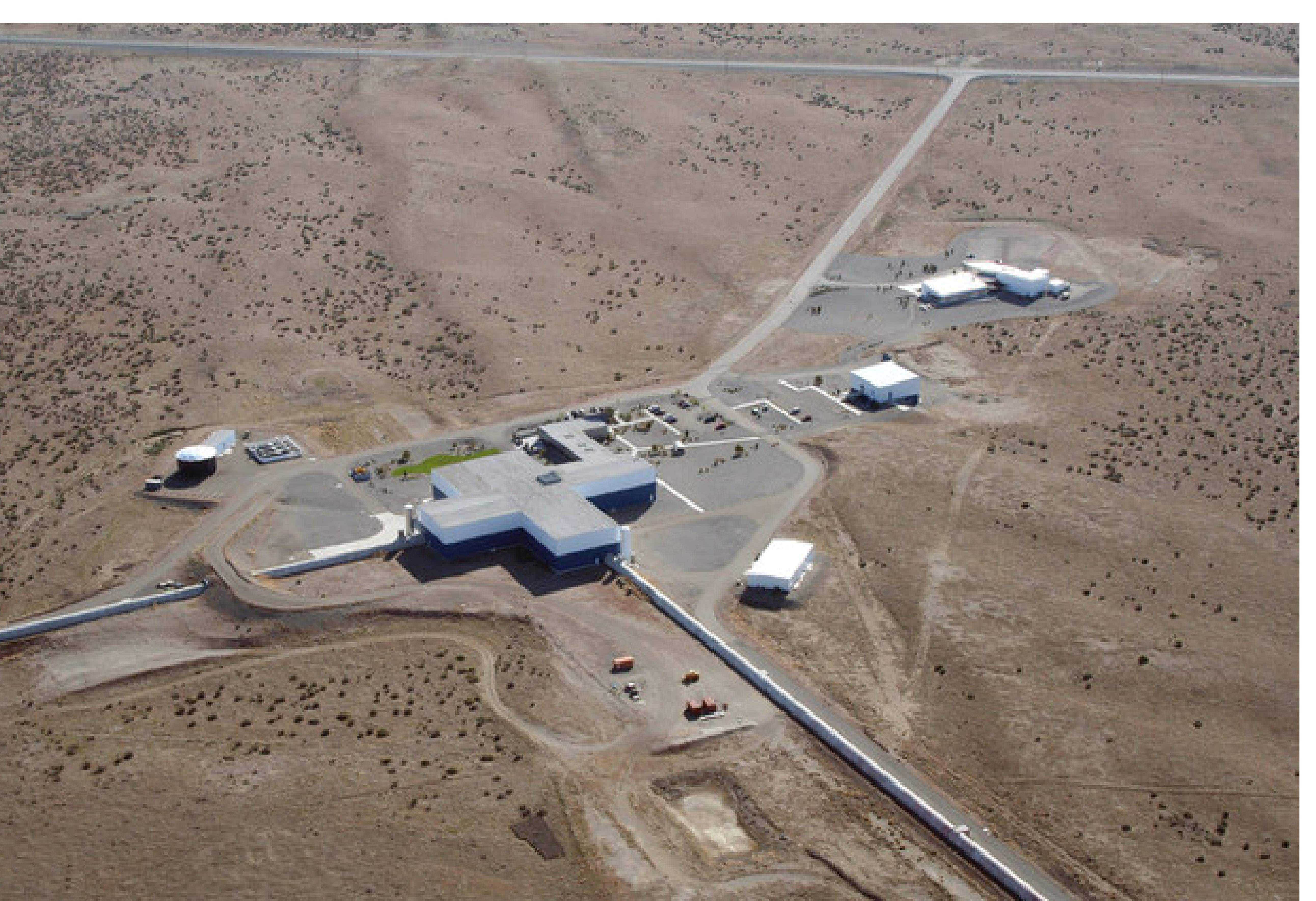} 
\caption{LIGO - Hanford, Washington DC. (Imagem disponível publicamente em https://www.ligo.caltech.edu/WA/image/ligo20150731d)} 
\label{ligo1} 
\end{minipage} \hfill 
\begin{minipage}[b]{0.45\linewidth} 
\includegraphics[width=\linewidth]{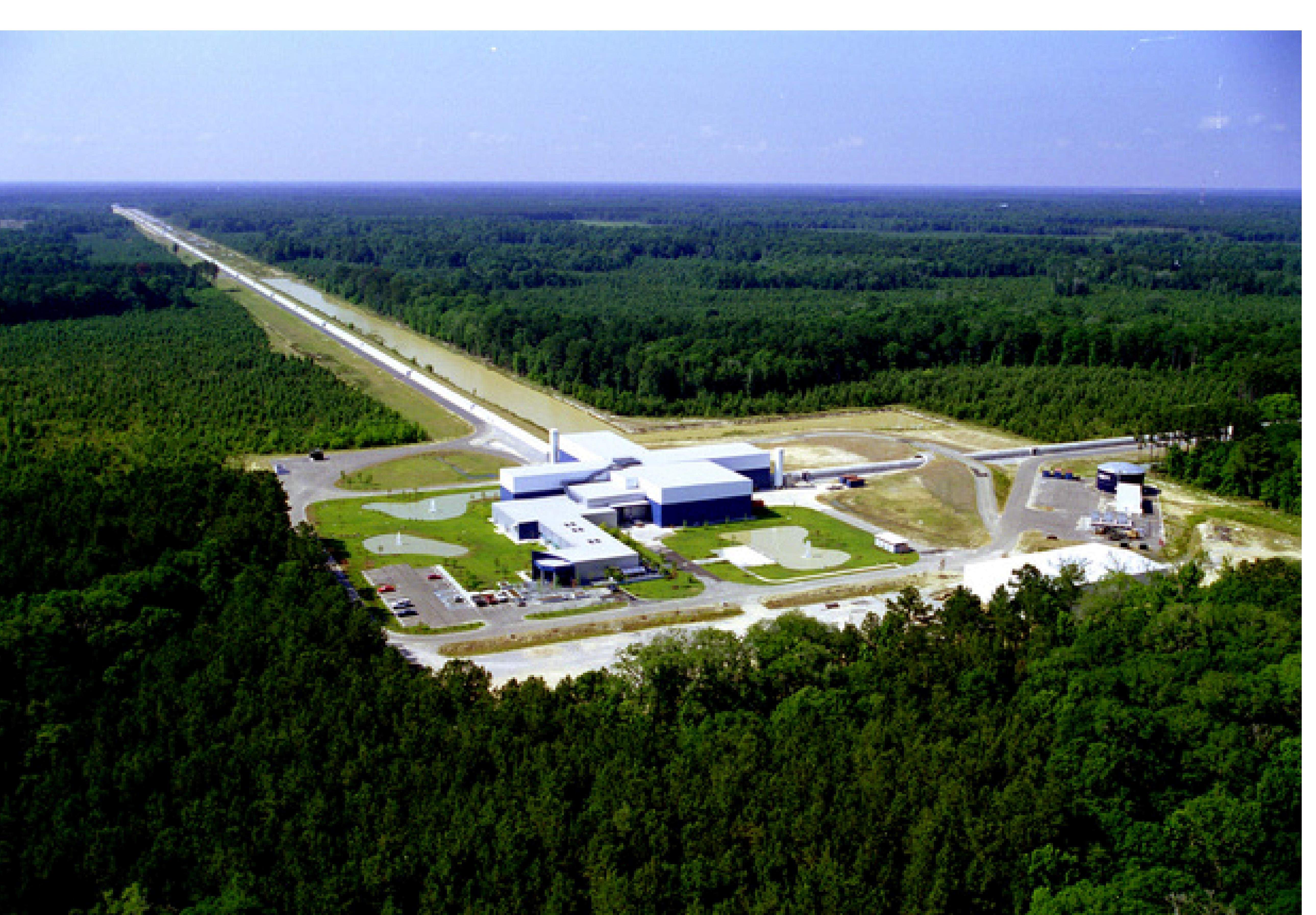} 
\caption{LIGO - Livingston, Louisiana. (Imagem disponível publicamente em https://www.ligo.caltech.edu/WA/image/ligo20150731c)} 
\label{ligo2} 
\end{minipage}
\end{figure*}

Todo o processo de captação das OG, desde a simulação do BBN até a prova concisa da existência das OG, ocorreu por técnicas de interferômetros de luz instalados em Hanford, Washington (cf. Fig.~\ref{ligo1}), e em Livingston, Louisiana (cf. Fig.~\ref{ligo2}), com um equipamento similar ao interferômetro de Michelson-Morley, produzido em 1887, com espelhos suspensos em sistemas a vácuo em braços de 3 a 4 km de extensão que acabavam por gerar padrões de interferência.

\begin{figure}[h]
\centering
\includegraphics[scale=0.6]{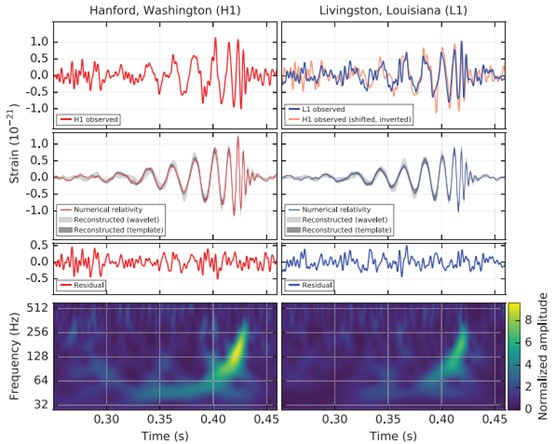}
\caption{Gráficos das amplitudes e frequências das OG detectadas pelo LIGO. (Figura retirada da Ref.~\cite{Abbott})}
\label{ligo_paper}
\end{figure}

As ondulações gravitacionais ao passarem pelo interferômetro mudam o padrão de interferência, informando que algo passou por aquela região. Como o sistema inteiro estava a vácuo e livre de qualquer influência externa, a mudança do padrão de interferência só poderia ocorrer em duas situações: ou pela mudança da fonte (feixe de luz), ou pela mudança dos braços, no quesito movimentação dos espelhos. Com essa logística, ficava mais claro a identificação das OG. Dados do artigo do LIGO~\cite{Abbott}, publicado em 11 de fevereiro de 2016, mostram que a primeira detecção, identificada por GW150914, chegou até o interferômetro com uma frequência de 35 a 250 Hz e uma amplitude muito baixa de aproximadamente $1,0 \cdot 10^{-21}$ m, como já era de se esperar, amplitudes muito baixas que inviabilizaram por muitos anos a sua detecção. Esse foi um ponto que comprovou que de fato se tratava de OG. A segunda detecção, identificada por GW151226, teve parâmetros semelhantes, com uma frequência que variava de 35 a 450 Hz e uma amplitude de onda de aproximadamente $3,4 \cdot 10^{-22}$ m e que segundo os autores da Ref.~\cite{Cattani}, “esses resultados confirmam que OG finalmente foram detectadas e que elas são geradas pela fusão do sistema BBN”~\footnote{Versão livre do original, \textit{"These results confirm that GW have finally been detected and that they are generated by merging BBH system".} (pag. 7)}. O gráfico da Fig.~\ref{ligo_paper}, demonstra as proporções de amplitudes e frequências captadas pelos dois laboratórios. Os detectores foram calibrados com frequências entre 35 a 350 Hz afim de filtrar o máximo as linhas espectrais observadas e suprimir as grandes flutuações não compatíveis com a calibração do equipamento. Segundo os dados, a onda gravitacional GW150914 chegou primeiro no detector de Louisiana (L1) e após, aproximadamente, alguns milisegundos, chegou ao detector de Hanford (H1), com comprimentos de onda na ordem de gerar os padrões de confirmação.
Os dois parâmetros (L1 e H1) são sobrepostos de forma invertida em termo de 94\%, como é possível ver no primeiro gráfico do lado direito, em L1, para que fosse subtraído as formas das ondas e filtrado os resíduos para melhor análise das curvas da onda, mostrados na terceira linha.
A última linha dos gráficos mostra as ondas sendo compatíveis com a calibração do detector, mantendo sua frequência aumentando com o tempo.

\subsection{Ondas gravitacionais direcionadas para o ensino médio}

Após apresentarmos de forma concisa o contexto histórico, físico e matemático do desenvolvimento das OG como teoria, já possuimos a capacidade de entender que tal comprovação é de fato um dos momentos mais importantes no avanço científico e que tem a capacidade de gerar inúmeros frutos de pesquisa e desenvolvimento no entendimento do Universo. Mas as perguntas básicas que fazemos são: de qual forma esse desenvolvimento e avanço pode influenciar o ensino de física nas escolas? Como o professor pode utilizar desses avanços para trazer luz à novas ideias em sala de aula seguindo os parâmetros curriculares escolares?

Antes de discutirmos uma solução para a aplicação do estudo das OG em sala de aula, levando o discente mais próximo da física de fronteira, é preciso entender em que cenário se encontra esse sistema de ensino de física e como o mesmo está sendo passado em sala de aula. Em seu trabalho~\cite{Nascimento}, Nascimento aponta uma ideia de qual cenário o ensino de física se apresenta, quando menciona que “pesquisas no mundo todo têm sugerido que o ensino de Física é via de regra, e salvo honrosas exceções, caótico, pouco frutífero e dicotomizado da realidade de professores e alunos”. 

Vale a pena discutir essa problemática de ensino, principalmente no quesito ensino de física, pois é na sala de aula que se encontra um dos maiores polos de divulgação científica que funciona basicamente como uma “faca de dois gumes”, onde, se utilizado de forma correta, pode-se preparar gerações e gerações a se desenvolverem cientificamente, ou, se utilizado de forma errada pode-se trazer uma visão totalmente desfocada do ensino de física, sendo que infelizmente é o que atualmente mais acontece. Também vale a pena ressaltar que existem uma série de impasses e variáveis que tornam o ensino de física algo totalmente mecânico e desvirtuado, no qual citaremos e discutiremos esses pontos com mais clareza nas próximas seções.

Já é de conhecimento que existe uma diferença entre o desenvolvimento da física e o ensino de física, que infelizmente em muitos casos, parecem não andar de mãos dadas. O que podemos observar é que os avanços da física estão bem à frente do ensino das suas implicações nas salas de aula, e isso é um grande problema, que nos motiva a propor soluções para resolver essa discrepância. O problema reflete principalmente na forma de tornar a ciência acessível à sociedade e consequentemente em desenvolver cientificamente esta sociedade, pois não tendo acesso ao desenvolvimento científico, não há condições de gerar mentes críticas à ciência. Podemos citar algumas afirmações que contribuem para a construção deste esteriótipo, como por exemplo: \textit{"alunos de escolas do ensino básico não possuem condições de aprender temas tão complexos"} ou \textit{"os alunos não irão precisar desse tipo de informação"} e isso acaba refletindo diretamente na ideia de que a “ciência é apenas para cientistas” e que “pessoas leigas não possuem condições de aprender algo um pouco mais avançado e complexo”, monopolizando o avanço científico e o excluindo do ensino de física que a muito se encontra desatualizado e repetitivo, dando a visão errada aos discentes de que a física apresentada em sala de aula é uma mera disciplina cuja funcionalidade é a aprovação perante aos sistemas escolares e vestibulares, em geral, sem nenhum raciocínio crítico, inovador e científico, como podemos enfatizar o que os autores da Ref.~\cite{Praxedes} mencionam em dizer que “as pesquisas relacionadas ao ensino de Física demonstram que o ensino atual tem assumido o caráter de preparação para a resolução de exercícios de vestibular”.

O ensino de física se encontra defasado no sentido de que boa parte do ano escolar do aluno, o mesmo só possui contato com a física clássica, enquanto observamos o desenvolvimento da física do século XXI, a física que os discentes possuem relação não passa do século XIX, como conclui Pereira e Aguiar~\cite{Pereira} em mencionar que “o ensino de física no nível médio tem se limitado, principalmente a temas da física clássica: mecânica, eletricidade e magnetismo, calor e óptica. Além disso, esse ensino se caracteriza, na maioria das vezes, por aulas teóricas e descritivas, distantes da realidade dos alunos”.

Terrazzan~\cite{Adolfo} também menciona sobre essa problemática dizendo que a prática escolar usual exclui tanto o nascimento quanto o entendimento da ciência, pois a grande concentração de tópicos se dá na física desenvolvida, aproximadamente, entre 1600 e 1850.

Vale ressaltar que a menção anterior sobre os discentes terem acesso apenas ao ensino de física clássica, em sua maioria, não desmerece em nenhum caso a sua importância no desenvolvimento científico do aluno. 

Segundo Nascimento~\cite{Nascimento}, “a física participa do desenvolvimento científico e tecnológico com importantes contribuições específicas, cujas consequências têm alcance econômico, social e político".  Esse sim, é o maior problema que se pode encontrar nas escolas atualmente, onde praticamente o aluno não é estimulado a pensar de forma crítica, a raciocinar de forma científica, mas apenas a “aprender” uma física repetitiva e que por sua vez acaba se tornando um fardo para o aluno. Em outras palavras, muitas vezes o conteúdo visto pelo discente em sala de aula não possui uma ligação com a sua realidade fora dela, sendo reafirmado por Moreira~\cite{Moreira} em mencionar que “o ensino da Física na educação contemporânea estimula a aprendizagem mecânica de conteúdos desatualizados”. 

Outro problema que será discutido na próxima subseção é a presença da Física Moderna e Contemporânea (FMC) em sala de aula, que é praticamente inexistente, mesmo estando presente nas bases curriculares nacionais, os PCN’s.  A explicação da escassez da presença da FMC nas escolas é bastante simples, a escola, em sua maioria, não atua mais na formação de novos conhecimentos para o desenvolvimento do aluno, acompanhando o desenvolvimento social, mas atua como um repetidor de ideias sem inovação e desenvolvimento científico do alunado, tendo como base regedora os próprios vestibulares, como menciona o educador Rubem Alves ao afirmar que são os vestibulares que definem as práticas escolares. 

Entendendo o real cenário no qual o ensino de física está imerso e todas as suas problemáticas, agora pode-se retornar a pergunta: Como o professor pode se utilizar do avanço científico das OG em sala de aula, dando ao aluno contato a uma física interessante e nova?

Assim como os cientistas e pesquisadores que se empolgaram com a comprovação das OG, assim também deveria se sentir um professor de física em saber que velhos conceitos, já presentes em sala de aula, podem ser renovados e novos apresentados aos seus alunos. Conceitos estes que possuem uma relação direta com o achado das OG, no qual mencionamos:

\begin{itemize}
\item O efeito ondulatório da luz, no quesito propagação construtiva e destrutiva de ondas, fazendo uma menção ao importante experimento de Michelson - Morley, datado como um dos mais memoráveis experimento da física; 

\item A abertura ao conhecimento da Relatividade Geral e Restrita de Einstein; 

\item A gravidade sobre corpos massivos e suas reais influências no tecido espaço-temporal, conceito este que fica vago no ensino comum abordado em sala de aula, resultando em que o discente ainda tem a ideia de espaço e tempo absoluto da mecânica de Newton; 

\item O próprio conhecimento sobre o espaço-tempo; 

\item Gerar então novas visões adentrando a um novo universo de conhecimento embasado pela física moderna.
\end{itemize}

São inúmeras as formas que a comprovação das OG podem ser apresentadas ao aluno pelo professor, principalmente ao mostrar que através desses novos conceitos pode-se entender mais a fundo o Universo em que conhecemos, mostrando além daquilo que somente as ondas eletromagnéticas podiam, trazendo então o aluno para uma área interessante da física, tornando o aprendizado mais significativo e assim tirando a visão errônea que a física só serve para uma determinada prova ou vestibular, mas que ela é um grande empreendimento humano, como abordado na Ref.~\cite{Rosella}.

Para melhorar o ensino da física em sala de aula é preciso de uma vigorosa renovação e conceitos interessantes como o de OG, principalmente com relação aos conteúdos que o aluno já possui contato. Esse é um passo fundamental para mostrar ao discente que o desenvolvimento da física ainda consegue surpreender e modificar o pensamento humano. Encontrando essa correlação, entre conceitos novos e aquilo que os discentes veem em sala, traz uma nova visão real da atuação da física, mostrando que a mesma não se limita apenas às atividades escolares, como muitos discente pensam. Essa correlação mostra ao aluno o que de fato a física é, uma ciência que tem como capacidade primordial mudar visões sobre o universo a sua volta se tornando aplicável e real. 

É compreensivo gerar a concepção de que independentemente das comprovações das OG de forma direta, os conceitos de ondulação, gravitação e os demais apresentados já seriam repassados para o aluno em sala de aula, mas é sabido entender que o professor, que de fato acredita no trabalho da ciência e que ela pode ser acessível a qualquer pessoa, pode ver esse momento da física contemporânea como um impulsionador para ministrar esses conteúdos de forma empolgante e realista, pois estão envolvidas em um recente avanço da física, compartilhando do seu entusiasmo pela ciência com seus alunos e assim gerando um aprendizado significativo. Todo esse contexto é endossado pela afirmação de Moreira~\cite{Moreira} em dizer que o “Ensino de Física não é uma questão de encher um cérebro de conhecimentos, mas de desenvolver esse cérebro em Física". 

A escola ainda se apresenta como o maior palco de divulgação científica e a ideia da inclusão não só de conceitos como as de OG mas de qualquer assunto referente a FMC, dá nova luz ao conhecimento do aluno, podendo até influenciar nas escolhas profissionais dos mesmos, como menciona os autores da Ref.~\cite{Renner} em enfatizar a presença de tais conteúdos em sala de aula podem sim influenciar jovens a procura de carreiras científicas, pois constituem, potencialmente, os futuros pesquisadores e professores de física. 

\subsubsection{Problemáticas do ensino de Física Moderna e Contemporânea nas escolas}

Abordaremos nesta subseção a problemática da inserção da FMC nas escolas, apontando possíveis motivos da não inclusão dessa temática e como resolver essa curricularização, afim de trazer uma maior compreensão e resolução do cenário em que o ensino de física está imerso.  

O cenário atual do ensino de física no País é de quase total estagnação quando olhamos para a grade curricular, onde ano após ano os mesmos assuntos e as mesmas aulas são ministradas, sem nenhum tipo de inovação do conhecimento apresentado em sala de aula, o que acaba por refletir de forma direta de como os discentes veem a disciplina. Sabendo disso, gera-se uma pergunta que acaba por embasar a nossa abordagem: Quais são as dificuldades encontradas no cenário do ensino de física que inviabilizam a manutenção do ensino e a fortificação do ensino de física moderna e contemporânea nas escolas?

Antes de responder a pergunta, precisa-se entender primordialmente qual a importância da presença da FMC nas escolas, pois sabendo da sua importância fica mais claro discutir sobre sua falta perante o ensino de física. A Ref.~\cite{Renner} ressalva sobre a inserção de tópicos relacionados à FMC no Ensino Médio ser de grande relevância, podendo proporcionar aos discentes o entendimento do mundo atual, tornado os conceitos mais significativos.

Logo, entende-se que a presença da FMC nas escolas traz ao aluno luz sobre novas fronteiras de conhecimento e cultura. Além disso, um contato direto com uma física mais interessante, despertando então no discente mais afinidade e consequentemente um melhor envolvimento com a disciplina. Agostin~\cite{Agostin} menciona essa importância da FMC para o discente ao dizer que a cultura científica deve fazer parte da cultura do educando. Isso  é possível se a inserção dos conceitos da Física Moderna e Contemporânea estiver associado a um processo de aplicação da cultura do educando.

Compreendendo a importância da FMC e tudo que ela pode proporcionar ao aluno, podemos agora apontar as principais causas da sua inconsistência em sala de aula e principalmente como tentar resolvê-las. O primeiro ponto crucial da problemática da presença da FMC em bases curriculares, vem dos próprios vestibulares que não possuem a estimativa de abordar tais temas em suas provas e bases, mesmo tendo o propósito em comum de abordar questões do cotidiano e de contextualizar problemas físicos, como o ENEM (Exame Nacional do Ensino Médio). Não existe nenhuma menção de física moderna, tanto analisando editais recentes em seus conteúdos programáticos disponibilizados pelo INEP (Instituto Nacional de Estudos e Pesquisas Educacionais Anísio Teixeira), quanto analisando as suas competências e habilidades da prova, o que se parece meio contraditório em se tratando do maior sistema de avaliação de conhecimento do país, pois a base regedora a educação brasileira, os  Parâmetros Curriculares Nacionais (PCN) menciona a importância do contato do aluno com os conteúdos atualizados e recentes, colocando assim o discente frente a inovações e avanços no conhecimento científico. Infelizmente, isso na prática, não é o que acontece.

Com base na visão apresentada, qual a relação das bases regedoras dos vestibulares na falha da inserção do ensino de FMC nas escolas? A resposta é simplória. Como mencionado anteriormente, as escolas, internamente, são direcionadas pelos vestibulares. Assim que é posto como conteúdo direcionado para tal exame de avaliação, as escolas, em sua maioria, adotam tal regimento focando na preparação e aprovação ao final da jornada escolar do aluno, não dando espaço para professores, na maioria das vezes, trabalharem assuntos fora dessa grade curricular, sendo esse então uma das maiores problemas encontrados que barram o avanço do conhecimento em sala de aula.

Outro ponto crucial na implementação da FMC no sistema escolar, vem muitas vezes por parte do professor, tanto no quesito de má preparação do professor, quanto no quesito ausência de formação na área, o que é mais comum do que se imagina, quanto na carência de sua formação nesses respectivos assuntos, como relatividade, ondas gravitacionais, física de partículas e muitas outras temáticas contemporâneas do ensino de física. Agostin~\cite{Agostin} menciona essa problemática em assegurar que alguns professores não se sentem seguros em lecionar conteúdos de Física Moderna e Contemporânea devido a uma formação inadequada.

Essa deficiência na formação acaba por se tornar uma grande barreira entre o professor e o aluno, sendo o professor uma espécie de canal de informação cujo fluxo é direcionado aos alunos; sem esse fluxo direcionado de forma correta, não há informação precisa. É lógico de mencionar que mesmo com essa problemática apresentada, a má formação docente, também entra em cena como um influente aspecto, o desestímulo do profissional professor que encara situações complicadas e exaustivas a cada dia de trabalho, com pouca valorização, fazendo muitas vezes com que o mesmo perca o interesse de renovar o seu conteúdo em sala de aula e em levar novas ideias para a discussão.

Apontando tais problemáticas de uma forma bem breve e direta, pode-se pensar em uma solução na tentativa de minimizar esse afastamento da FMC das escolas e preparar o campo escolar para a futura incursão da FMC nos currículos escolares. É indiscutível o fato de que é necessário uma renovação rápida no ensino de física a fim de trazer um aprendizado mais significativo, saindo da inércia em que a física entrou dentro da sala de aula. A maior renovação deve vir por parte, principalmente, dos sistemas avaliativos que dão ao aluno ingresso nas Universidades, pois sendo eles, em sua maioria, os regentes daquilo que é apresentado nas escolas, fazendo valer o que os PCN defendem e adotando mais conceitos de FMC em seus testes, no objetivo de avaliar não somente a capacidade conteudista dos alunos, como tem feito atualmente, mas sim de avaliar como anda o aluno perante aos avanços científicos atuais, o avaliando pela capacidade de se contextualizar perante tais avanços. Isso traria um aprendizado mais significativo e consequentemente mais interesse dos alunos nessa área científica. No mesmo segmento, os vestibulares adotando temas como os de FMC em seus testes, as escolas acabariam por tentar se adequar a essa nova investida dos vestibulares e passariam a dar mais espaço ao ensino de FMC em seus estabelecimentos, visando uma melhor preparação dos seus alunos. Consequentemente, com a mudança das temáticas dos vestibulares e as escolas se adaptando a elas, o professor teria uma maior liberdade para compartilhar com seus alunos tais avanços científicos em sala de aula e investindo mais em sua preparação pessoal e profissional, gerando então um efeito em cadeia.

Claro que toda essa proposta iria levar algum tempo para ter mais clareza na sua atuação, mesmo com inúmeras pesquisas que buscam ajudar na renovação do ensino como apresenta Menezes~\cite{Menezes} em dizer que, 

\begin{quote}
É claro que precisa ser cautelosa a sinalização para a inclusão desses novos conteúdos, seja pelos desafios didáticos que implica, encontrando professores despreparados e os textos escolares desguarnecidos, seja porque as próprias universidades, ainda por algum tempo, continuarão a solicitar os velhos conteúdos em seus vestibulares. Será preciso algum tempo para que a mensagem seja, primeiro, compreendida e, mais tarde, aceita. (pág. 7)
\end{quote}

É indiscutível a presença da FMC nas escolas, pois prepara seus alunos para a vida, a sociedade e principalmente para os avanços tecnológicos. Com base na exposição dessa problemática, vamos apresentar na seção seguinte uma proposta metodológica para a inserção da FMC nos conteúdos de física, tomando como base o conteúdo de OG.


\section{Metodologia e aplicação}

Para o desenvolvimento da pesquisa usamos uma metodologia de caráter empírico e avaliativo que foi empregada na obtenção de dados necessários para a discussão dos objetivos do trabalho. Com isso, estabelecemos 3 estágios para o seu desenvolvimento:

\begin{enumerate}
\item Apresentação em sala de aula do tema ondas gravitacionais e a discussão sobre sua importância e envolvimento de tal avanço com o currículo escolar dos estudantes presentes;

\item Apresentação do experimento de Michelson-Morley caseiro em sala de aula;

\item Avaliação/Questionário de caráter objetivo e com a intenção de captar dados daquilo que foi compreendido pelos alunos presentes, que responderam a questões do tipo:

\begin{itemize}
\item Você se acha capaz de aprender assuntos que advenham da física moderna como ondas gravitacionais?

\item O que você conseguiu absorver/aprender da aula de introdução sobre ondas gravitacionais?

\item Acredita que o assunto de ondas gravitacionais abordado em sala de aula tem alguma relação direta com conteúdos que você já viu nos anos anteriores?

\item O que você acha da presença de conteúdos da física moderna e contemporânea acrescidos ao currículo escolar? Pode trazer algum benefício ao aprendizado do aluno? Se sim, especifique.

\item A apresentação do experimento de Michelson-Morley melhorou o seu entendimento sobre as ondas gravitacionais e sua detecção?

\item Encontrou alguma dificuldade durante toda a aula de apresentação e demonstração do experimento?
\end{itemize}
\end{enumerate}

\subsection{Construção do interferômetro de Michelson-Morley}

Como uma das propostas desse trabalho, a construção e apresentação do experimento de Michelson - Morley é usado para dar um embasamento da ideia de funcionamento dos interferômetros responsáveis pela comprovação das ondas gravitacionais, pelo LIGO, dando ao aluno um contato mais visual e real do seu funcionamento, já que o assunto de ondas gravitacionais é algo um pouco abstrato e que precisa de mais aprofundamento. A construção do experimento tem também a importante missão da inserção da física experimental em sala de aula, mostrando que a sua presença sempre foi uma excelente ferramenta para sair da inércia do ensino de física e trazer mais luz ao conhecimento apresentado, despertando sempre a curiosidade e o lado investigativo do aluno. 

O experimento do interferômetro de Michelson - Morley é um dos mais importantes experimentos já realizados na física, sendo responsável basicamente pelo transpassar de uma física clássica para uma física moderna, trazendo conceitos importantes e quebras de paradigmas que influenciaram diretamente e completamente os passos seguintes do conhecimento físico científico. O seu principal motivo de formulação foi a tentativa de entender mais a fundo o funcionamento da luz e as suas particularidades, já que a comunidade científica aplaudia e se vangloriava com os avanços de Maxwell principalmente pela dada unificação entre a eletricidade e o magnetismo. 

O experimento, fundamentado por Albert Michelson e Edward Morley, em 1887, acabou por ser conhecido por um fator fundamental na movimentação da sociedade entre a física clássica e a física moderna, acerca do entendimento do Éter luminífero, propriedade física criada a fim de fundamentar em qual meio a luz deveria se propagar, em analogia com o som, além de fundamentar muitos ideais newtonianos aqui já comentados. Como é de conhecimento, o resultado do experimento foi negativo, descartando a ideia do Éter como propriedade presente no que chamamos de espaço-tempo; um outro resultado do experimento foi a confirmação, já predita por Maxwell, da constância da velocidade luz.

É de imaginar o quanto esses resultados influenciaram no entendimento “progressivo” da física, mudando as concepções físicas estabelecidas, dando espaço para o desenvolvimento da relatividade. Einstein menciona que para desenvolver sua teoria da Relatividade não teve a necessidade de usar nenhum dado advindo de Michelson-Morley, mas que se o soubesse na época da formulação, com certeza iria mencionar o grande feito dos dois cientistas.

A ideia de interferômetro volta à tona tanto na década de 1970, quando foi posto o início para o desenvolvimento tecnológico de equipamentos capazes de detectar as ondas gravitacionais, quanto agora no século XXI, em 2015, quanto a primeira onda gravitacional foi detectada diretamente. Muitos equipamentos foram desenvolvidos com o decorrer dos anos, em diversos países, inclusive no Brasil, com o experimento Mário Schenberg~\cite{Souza} que atualmente está em posse do INPE (Instituto Nacional de Pesquisas Espaciais), com a finalidade de detecção das tão esperadas ondas de Einstein. Mas, entre todas as tecnologias e equipamentos desenvolvidos, os interferômetros tomaram a posse do equipamento capaz de realizar tal feito. Daí a importância que esse trabalho teve em levar um protótipo do interferômetro de Michelson-Morley para a sala de aula, em decorrência da sua importância atual e na tentativa de elucidar as ideias dos alunos sobre a formalização das ondas gravitacionais.

O experimento em sala de aula é desafiador e complicado de ser elaborado, pois possui uma série de variáveis que precisam ser minimizadas, como por exemplo as vibrações presentes no ambiente de execução que podem gerar mudanças no padrão de interferência das ondas, para então ter-se a efetivação do experimento com a geração do seu padrão de interferência, sendo esse um dos maiores empecilhos da sua utilização em sala de aula por parte dos professores de física. Devido a alta sensibilidade do experimento, utilizamos materiais simples em sua fabricação como:

\begin{itemize}
\item Base metálica de 70,0 cm de comprimento por 50,0 cm de largura;

\item 2 espelhos refletores, cada um com 9,0 cm de lado;

\item 1 divisor de feixes de vidro cúbico espesso de 9 cm de lado e 0,50 cm de espessura (podendo ser utilizado outros materiais como acrílico espesso, mas no caso do experimento para a execução desse trabalho foi usado o vidro);

\item 1 laser pointer de 8000 mW de potência, 532 nm de comprimento de onda, utilizado como fonte de luz; 

\item 1 anteparo branco de papel.
\end{itemize}

\begin{figure}[h]
\centering
\includegraphics[scale=0.80]{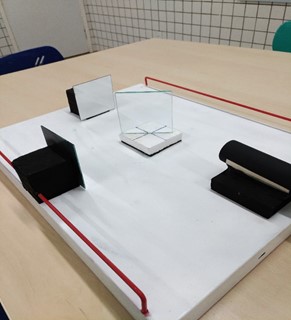}
\caption{Interferômetro de Michelson-Morley construído com materiais de baixo custo.}
\label{aparato1}
\end{figure}

Cada componente do aparato experimental, desde os espelhos, divisor do feixe e a fonte de luz foram montados sobre bases. Os espelhos e a fonte de luz foram colocados em uma base de madeira, e para a base do divisor de feixe foi utilizado o material de isopor a fim de minimizar as vibrações do meio. Com o mesmo motivo de minimizar as  vibrações e entendendo que a fonte é um dos mecanismos mais importantes para a execução do experimento, a fonte de luz foi coberta por um material esponjoso e colocado sobre uma estrutura previamente definida para auxiliar no direcionamento do feixe de luz (cf. Fig.~\ref{aparato1}). 

Outros materiais também poderiam ter sido utilizados para complementar o experimento, como por exemplo, a massa de modelar. A mesma poderia ser colocada abaixo da estrutura dos espelhos e também da base que sustenta os componentes com a finalidade de minimizar o efeito de vibração sobre o experimento. Também pode ser usado uma lente focalizadora convergente na saída dos feixes para melhor definição do padrão de interferência. Ambos os materiais não foram necessários para a execução positiva do experimento, sendo, portanto, possível fazer a observação dos padrões de interferência de forma nítida e clara sem o auxílio dos mesmos. Ressaltamos que um dos fatores importantes que levou ao bom funcionamento do experimento foi justamente a potência do laser utilizado. Foram realizados alguns testes com outros tipos de lasers de potências menores e os resultados foram insatisfatórios devido a pouca nitidez e as vezes nenhum tipo de padrão de interferência possível de se visualizar. 

\begin{figure}[h]
\centering
\includegraphics[scale=0.65]{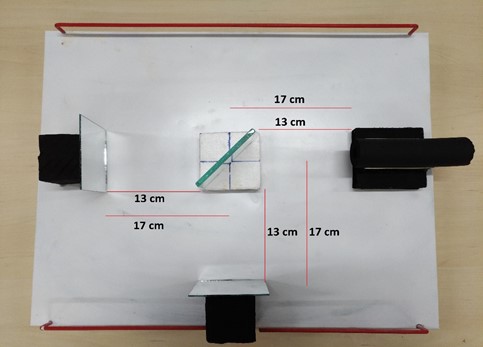}
\caption{Medidas para a construção do interferômetro de Michelson-Morley caseiro.}
\label{aparato2}
\end{figure}

Observamos que a maior dificuldade encontrada na montagem do experimento se deu na calibragem do mesmo. Para o seu bom funcionamento, todas as medidas de separação entre divisor-espelhos e divisor-fonte foram feitas cuidadosamente para que os feixes fizessem os mesmos caminhos ópticos e gerassem no anteparo um padrão contínuo e observável. Como podemos observar na Fig.~\ref{aparato2} a separação entre o centro do divisor e os espelhos foi de 17,0 cm e a separação dos espelhos até a base de sustentação do divisor foi de 13,0 cm, assegurando assim que o feixe percorresse o mesmo caminho ótico tanto para o espelho 1, quanto para o espelho 2. A distância entre a base da fonte de luz e a base do divisor foi de 13,0 cm e de 17,0 cm até o centro da base do divisor de feixes, mantendo uma regularidade no caminho ótico do feixe desde a sua saída da fonte até a sua propagação no anteparo. Essa regularidade de caminho ótico se mostrou importantíssimo para a execução de forma positiva do experimento, sendo que em testes anteriores, os padrões de interferência não se efetivam ou se apresentavam com pouca nitidez e difícil visualização.

\begin{figure}[h]
\centering
\includegraphics[scale=0.08]{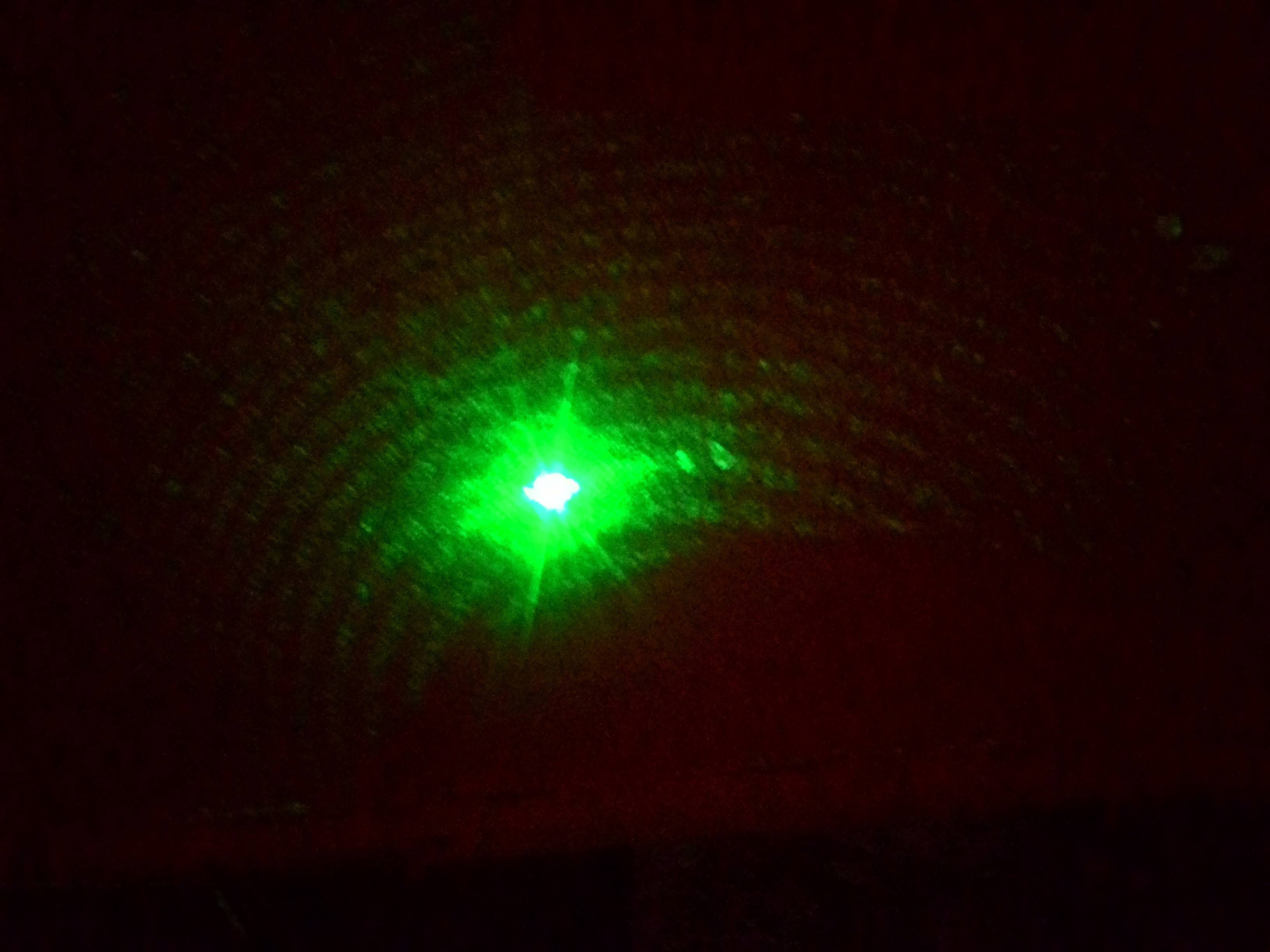}
\caption{Padrão de interferência encontrado em testes do interferômetro caseiro.}
\label{franjas}
\end{figure}

Depois de regular o caminho óptico e a calibração estrutural do experimento gerou-se então padrões limpos e nítidos de interferência como também a sua observação de variação do padrão quando movimentado um dos espelhos, variando o caminho ótico e simbolizando como paralelamente ocorreu no interferômetro do LIGO quando a onda gravitacional se propagou pelos espelhos, mudando-os e alterando o caminho ótico, gerando padrões também alterados. A Fig.~\ref{franjas} mostra o resultado dos padrões de interferências encontrados com o interferômetro caseiro.

Para findar a parte de especificações do experimento, todo o aparato proposto acima, foi posto sobre uma base de metal acolchoada embaixo para minimizar mais ainda as vibrações do meio.

Toda a construção do projeto do interferômetro caseiro saiu em torno de 20 reais, sendo que a maioria dos componentes utilizados foram reaproveitados e podem ser encontrados com muita facilidade. Isso mostra que é um experimento de alta excelência, com importâncias significativas, mas com muito baixo custo, sendo acessível a qualquer profissional da área que tenha a idealização de melhorar o seu ensino em sala de aula, levando os alunos a terem contato com algo mais palpável e visual. O maior problema que se encontra com esse tipo de experimento é a sua calibração e utilização, pois como se trata de um experimento muito sensível, o ambiente em si acaba interferindo muito na sua execução, o que dificulta a realização desse experimento por parte dos professores que muitas vezes não tem tempo algum para utilizarem em sala de aula. 

\subsection{Projeto em sala de aula}

O segundo passo de execução e conclusão dessa pesquisa foi ministrar os conceitos teóricos de OG em sala de aula, apresentando o interferômetro caseiro e dando aos alunos conhecimentos renovados e interessantes, principalmente fazendo uma correlação com os conteúdos que os mesmos possuem em sala de aula. 

O local escolhido para a prática do projeto foi uma escola particular da cidade de Picos (PI), sendo aplicado em turmas de 2º e 3º ano, justamente pela capacidade de conteúdos já vistos pelos mesmos e envolvendo alunos com diferentes níveis de conhecimento de física, ficando então mais viável apresentar o projeto. O total de alunos participantes do projeto foi de 128 alunos, separados em 3 salas diferentes, que tiveram contato tanto com a ideia de ondas gravitacionais como também com o funcionamento do interferômetro, gerando ao final uma série de questionamentos e interesses. O projeto foi apresentado em três etapas:

\begin{enumerate}
\item A primeira etapa se deu na apresentação do embasamento teórico de ondas gravitacionais. Discussão sobre a importância dessa comprovação na física moderna e a sua relação com os conteúdos já vistos pelos alunos, trazendo o ideal apresentado para algo mais próximo da convivência dos alunos. Uma aula expositiva de 30 a 40 minutos, cujo principal objetivo era fornece conhecimentos necessários e básicos para o entendimento prévio das ondas gravitacionais e consequentemente do funcionamento do interferômetro.

\item A segunda etapa se efetuou com a apresentação do interferômetro caseiro para os alunos a fim de fundamentar o funcionamento dos laboratórios do LIGO.

\item A terceira e última etapa foi formulada com a aplicação de um questionário simples e objetivo com os questionamentos apresentados no início desta seção.
\end{enumerate}

Um importante relato que devemos mencionar, como fruto da aplicação do projeto em sala de aula, foi o interesse que os alunos tiveram ao se deparem com uma física totalmente fora da convivência deles, mesmo que aparentasse ser um conteúdo bem complexo e dificultoso, por um momento os alunos que tiveram contato com o projeto, deixaram um pouco de lado essa barreira que muitas das vezes parece ser intransponível quando fala-se em ensino de física e se atentaram a novidade apresentada ali, levantando questionamentos de suma importância para o acréscimo desta pesquisa. Isso mostra que de fato o ensino de física nas escolas precisa urgentemente de uma renovação curricular, como já mencionado anteriormente.

\begin{figure*}[!ht]
\begin{minipage}[b]{0.47\linewidth} 
\includegraphics[width=\linewidth]{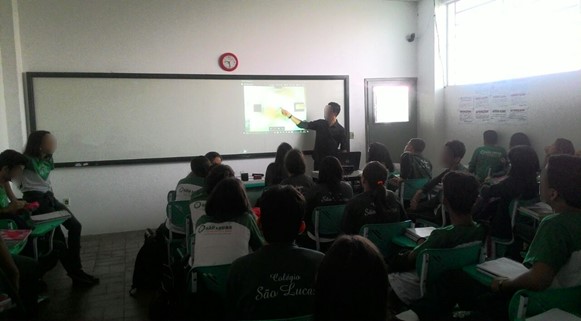} 
\caption{Aula teórica sobre ondas gravitacionais.} 
\label{aula1} 
\end{minipage} \hfill 
\begin{minipage}[b]{0.47\linewidth} 
\includegraphics[width=\linewidth]{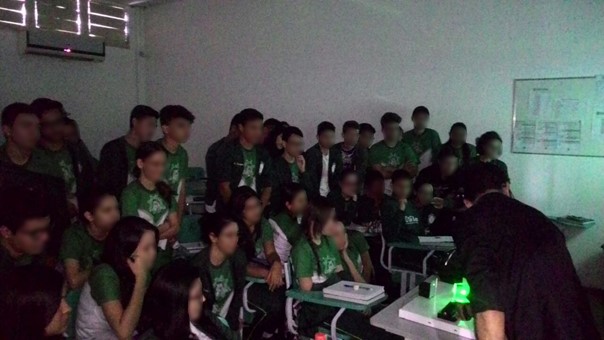} 
\caption{Apresentação do protótipo do interferômetro caseiro.} 
\label{aula2} 
\end{minipage}
\end{figure*}

\begin{figure*}[!ht]
\begin{minipage}[b]{0.47\linewidth} 
\includegraphics[width=\linewidth]{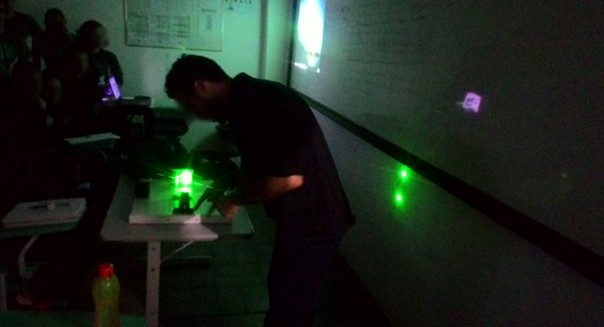} 
\caption{Amostra Experimental do Interferômetro.} 
\label{aula3} 
\end{minipage} \hfill 
\begin{minipage}[b]{0.43\linewidth} 
\includegraphics[width=\linewidth]{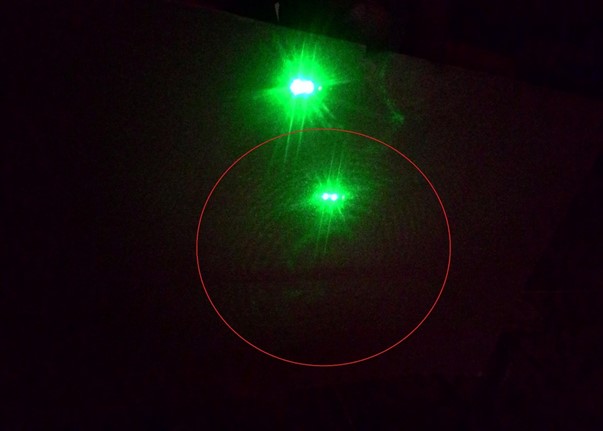} 
\caption{Franjas de interferência encontradas na apresentação do projeto.} 
\label{aula4} 
\end{minipage}
\end{figure*}

Ressaltamos que alguns problemas foram encontrados quanto a apresentação em sala de aula, pois sendo um experimento sensível e que necessita de um local com baixa luminosidade para assim concretizar a visualização das franjas de interferência, algumas das salas não contavam com essa baixa luminosidade, o que atrapalhou um pouco a execução em sala de aula, mas na tentativa de suprir essa falta, testes com o interferômetro foram feitos anteriormente à apresentação e por seguinte mostrados aos alunos presentes em sala. Tal empecilho não influenciou em nenhum quesito na opinião dos alunos dado nos questionários apresentados. Nas Figs.~\ref{aula1}-\ref{aula4} pode-se acompanhar sobre a apresentação do projeto em sala de aula.

\section{Resultados e discussão}

Após todo um processo de desenvolvimento do protótipo do interferômetro caseiro de Michelson-Morley e apresentação do projeto em sala de aula, onde levamos para a sala o desenvolvimento da física moderna através da temática de ondas gravitacionais, coletamos dados através de um questionário para discutirmos acerca das ideias debatidas neste trabalho e que nessa seção discutiremos sobre a presença das ondas gravitacionais no ensino de física e quais os seus benefícios perante aos discentes.

Os alunos envolvidos no processo de coleta de dados se depararam, após uma breve discussão e apresentação do conceito de ondas gravitacionais, com questões que acabaram por fazê-los raciocinar e darem respostas que mostram de fato como o aluno vê o ensino de física perante o mesmo, colocando declarações que aqui serão mencionadas a fim de embasar mais ainda as ideias apresentadas neste trabalho. Um dos questionamentos fundamentais dessa pesquisa, desde o início de sua formulação, foi de saber se os alunos que estão em séries do Ensino Médio, com currículos escolares preenchidos com o conteúdo da física convencional, poderiam ter certa capacidade de se familiarizarem com conceitos de física moderna e contemporânea em sua trajetória de conhecimento, pois não adianta fazer uma renovação do currículo escolar quanto ao ensino de física, acrescentando ideais de física moderna e contemporânea, que já entendemos ser de grande importância para o desenvolvimento científico dos alunos, sem ao menos entender se os mesmos são capazes de lhe dar com essas novas ideias e trabalhar em cima das mesmas. Com base no que foi apresentado em sala de aula e com os dados coletados a partir dos alunos, que foram instruídos a serem os mais verdadeiros e coerentes em suas respostas, observou-se que dentre os 128 entrevistados, 115 acreditam que são capazes de aprender e lhe dar com conteúdo de física moderna e contemporânea, sobre a prerrogativa de que esse tipo de conteúdo gera mais \textit{“interesse, estímulo e curiosidade para aprender física”} nas palavras de um dos alunos entrevistados, fazendo jus ao que foi mencionado anteriormente. 

\begin{figure}[h]
\centering
\includegraphics[scale=0.4]{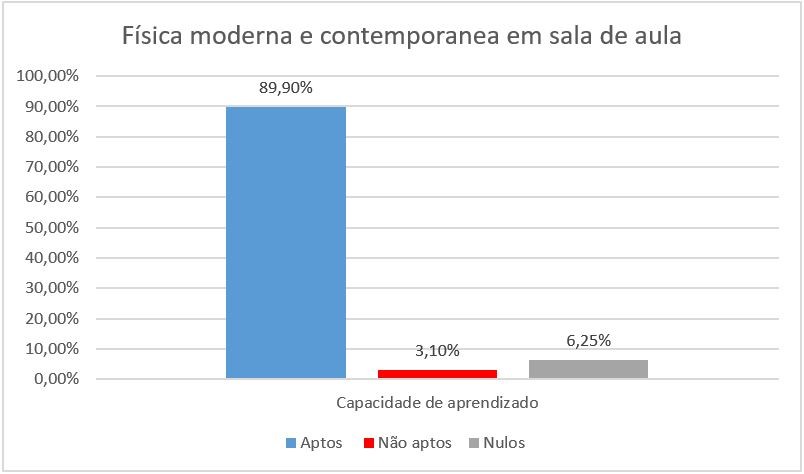}
\caption{No gráfico estão relacionados, em termos porcentuais, a capacidade de aprendizado do aluno em temas de Física Moderna e Contemporânea.}
\label{g1}
\end{figure}

Dos 128 entrevistados, apenas 4 acreditam que não são capazes de lhe dar com os ideais da física moderna e contemporânea com a prerrogativa de acharem complexo e distante daquilo que conhecem, mantendo a barreira da dificuldade presente. Como podemos observar no gráfico da Fig.~\ref{g1}, quase 90\% dos entrevistados acreditam ser apto em discutir conceito físicos mais modernos, enquanto 3,1\% não acreditam nessa possibilidade. Já 6,25\% dos entrevistados se recusaram a responder esse questionamento.

Esses dados são de grande importância para a inclusão de conceitos de física moderna e contemporânea em sala de aula, servindo como indicativo para a sua presença perante os discentes, quebrando o estereótipo de que pessoas leigas, alunos leigos a esse tipo de desenvolvimento científico não possuem capacidade de assimilá-lo e compreendê-lo mesmo que de forma superficial.

\begin{figure}[h]
\centering
\includegraphics[scale=0.40]{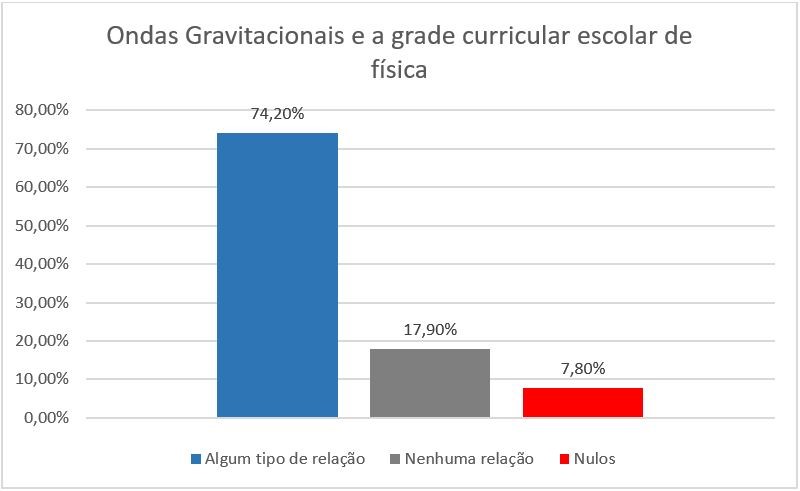}
\caption{No gráfico mostramos a correlação, em termos porcentuais, entre o conceito de ondas gravitacionais e assuntos já vistos em sala de aula pelos alunos.}
\label{g2}
\end{figure}

Outro ponto chave deste trabalho era mostrar que o ideal de ondas gravitacionais está presente naquilo que os alunos já conhecem em sala de aula, mostrando que  existe uma forma do professor, preocupado com o desenvolvimento científico dos seus alunos e não apenas nas suas aprovações, introduzir avanços científicos como os de ondas gravitacionais no currículo escolar, se entusiasmando juntamente com seus alunos com os avanços da ciência. Esse é outro indicativo bem defendido nesse trabalho, dando uma abertura a mais para a inclusão desses ideais. Com base no que foi debatido em aula e nos dados colhidos, observou-se que cerca de 74,2\% dos entrevistados conseguem ver algum tipo de relação com aquilo que aprendem ou aprenderam em sala de aula com o avanço da comprovação direta das ondas gravitacionais como podemos observar no gráfico da Fig.~\ref{g2}.

Considerando ainda o gráfico da Fig.~\ref{g2}, quase 18\% dos avaliados mencionaram que não conseguem fazer uma correlação com um avanço recente da observação das ondas gravitacionais e os conteúdos vistos em sala. Para critério de complementação, quase 8\% dos entrevistados declinaram em responder esse questionamento. A conclusão que se pode chegar com esse segundo dado é que existem formas tanto do professor quanto do aluno em desenvolverem ideias de física moderna e contemporânea sem sair da grade curricular, servindo como porta de entrada para futura inserção da FMC em sala de aula.

\begin{figure}[h]
\centering
\includegraphics[scale=0.40]{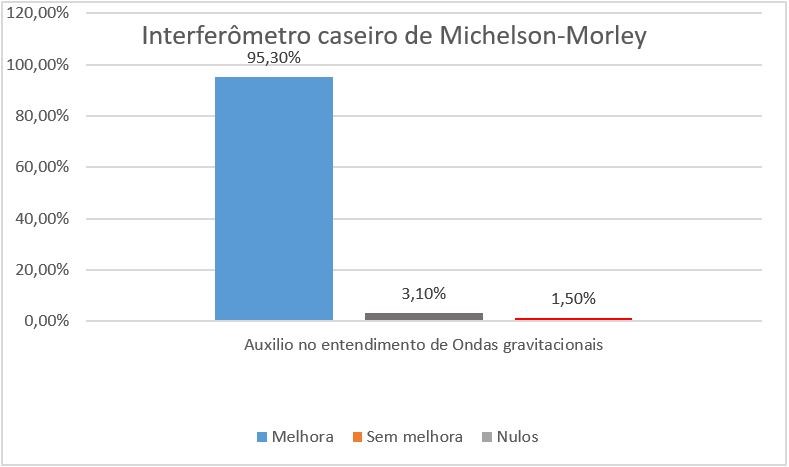}
\caption{No gráfico mostramos, em termos porcentuais, a aplicação do experimento de interferômetro de Michelson-Morley como facilitador no entendimento sobre as ondas gravitacionais.}
\label{g3}
\end{figure}

Seguindo com a análise dos dados, como segunda etapa do projeto, foi apresentado o protótipo do experimento de interferômetro de Michelson-Morley como uma analogia ao funcionamento os laboratórios do LIGO, a fim de trazer o avanço gravitacional para mais próximo do aluno. Mesmo parecendo muito distante da realidade, mas com base no experimento apresentado em sala de aula, dos 128 alunos entrevistados, quase a sua totalidade, 122 alunos (95,30\%), viram que a apresentação do experimento de Michelson-Morley  ajudou consideravelmente no entendimento mais a fundo do desenvolvimento e processo da comprovação das ondas gravitacionais como podemos observar no gráfico da Fig.~\ref{g3}. Para questão de complementação, apenas 3,10\% dos alunos não observaram melhoria no entendimento do assunto de ondas gravitacionais com o auxílio do experimento, enquanto 1,5\% não quiseram opinar sobre esse questionamento. Esse resultado corrobora com o que já foi mencionado em seções anteriores deste trabalho e em outras pesquisas que defendem de forma indiscutível a presença de experimentos em sala de aula para enriquecer o ensino de física.

\begin{figure}[h]
\centering
\includegraphics[scale=0.4]{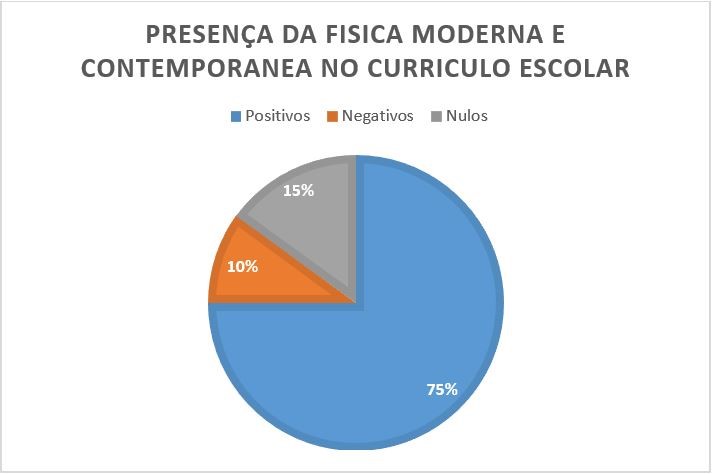}
\caption{No gráfico mostramos, em termos porcentuais, a opinião dos discentes acerca da presença da física moderna no currículo escolar.}
\label{g4}
\end{figure}

Discutiremos agora o que os avaliados acham do acréscimo de assuntos de física moderna e contemporânea à sua grade curricular escolar e como ela pode ser útil no seu desenvolvimento. De antemão, pode-se imaginar uma certa relutância da parte dos alunos em aumentar a sua carga horária de estudos em física ou modificá-la, acrescentando assuntos que os mesmos mal possuem contato, mas com base nos dados obtidos na pesquisa, de uma forma contrária do que se imaginava, houve uma aceitabilidade positiva por parte dos discentes nessa questão de acréscimo de conhecimento, mostrando o interesse pelos estudos e contatos com a física contemporânea. Dos 128 dados obtidos, cerca de 75\% acreditam e apoiam a inserção desse tipo de abordagem física em sala de aula, como podemos observar no gráfico da Fig.~\ref{g4}. Esse é um dado bem significativo e positivo para a presença da FMC em sala de aula, levantando mais uma vez questões da mesmice e ultrapassada física apresentada em ambiente escolar com uma certa desaprovação expressada pelos próprios discentes avaliados.

Vale mencionar alguns dos comentários dos alunos avaliados que se expressavam em dizer que acham \textit{“muito importante”} e acreditam na ideia do incremento da física moderna e contemporânea em sala e que estão \textit{“presos a descobertas físicas passadas e acabam alienados a questões relacionadas a física moderna, que caso, estudada em sala de aula, traria muitos benefícios”}. Outro aluno menciona que os conteúdos da escola em ensino de física são \textit{“retrógrados e que necessitam de uma renovação”}, fazendo jus à proposta do trabalho. Outros avaliados mencionam que \textit{“o estudo da física moderna é imprescindível para que os estudantes sejam capazes, além de compreender fenômenos, de ter mais visibilidade sobre o que o estudo da física pode proporcionar de avanços nas mais diversas áreas”}, além também de mencionar a correlação entre o apresentado no projeto e aquilo que os mesmos tem contato em sala de aula em mencionar que \textit{“é uma ideia bastante relacionada a outras, como gravitação e óptica, o que gera uma maior absorção desses conteúdos”}. Todas as declarações aqui mencionadas reforça o que foi defendido no trabalho e mostrando que de fato é necessário uma renovação do ensino de física, dando aos alunos contato com as evoluções e revoluções científicas, mas que também os próprios discentes aparentam desejar essa mudança e renovação.

Ainda com base no gráfico da Fig.~\ref{g4}, apenas 10\% acreditam não achar necessária essa mudança por defenderem a ideia de que tal temática deve ser direcionada apenas a aqueles que cursam ensino superior na disciplina, ou seja, em física, mantendo o estereótipo inicial e 15\% dos entrevistados não quiseram participar dessa avaliação.

\begin{figure}[h]
\centering
\includegraphics[scale=0.4]{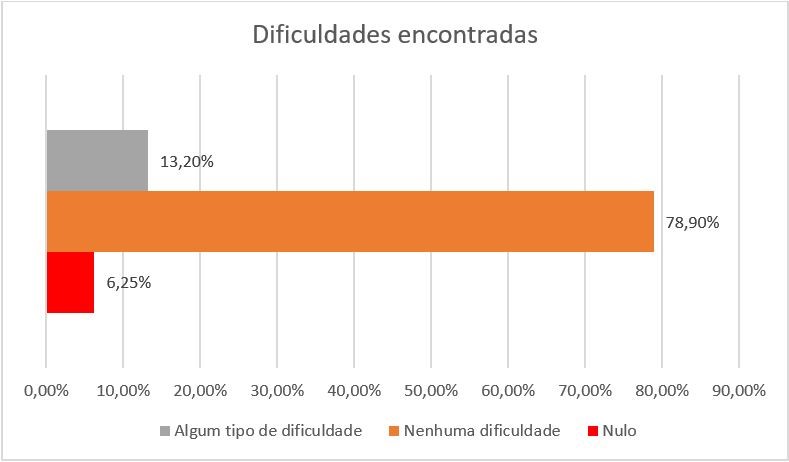}
\caption{No gráfico mostramos, em termos porcentuais, a análise de dificuldades de compreensão por parte dos alunos com o tema apresentado.}
\label{g5}
\end{figure}

Para findar a análise e discussão de dados, durante todo o processo de apresentação do projeto em ambiente escolar, apenas 13,2\% dos 128 alunos avaliados encontraram algum tipo de dificuldade, seja ela experimental ou até mesmo de entendimento e raciocínio do conteúdo de ondas gravitacionais. Em contrapartida, quase 80\% dos avaliados não encontraram uma dificuldade para expressar como podemos observar no gráfico da Fig.~\ref{g5}. Apenas 6,25\% não participaram de tal questionamento.

Após todo o processo de análise e discussão dos dados obtidos, temos bons indícios para implementar a presença da FMC em sala de aula, principalmente pelas opiniões por parte dos alunos que observaram essa necessidade. A física moderna e contemporânea tem muito a acrescentar no crescimento científico dos alunos, os tornando mais abertos ao desenvolvimento e gerando mais qualidade no ensino e no aprendizado, como os próprios entrevistados abordaram durante a aplicação do projeto. Talvez o “efeito dominó” volte a ser aplicado nesse ponto, onde a partir do momento em que os alunos tiverem interesse nessa área, os professores talvez se preocupem mais em se prepararem para apresentar e levar a seus alunos um conhecimento mais abrangente, onde futuramente sendo temas de sistemas avaliativos e como consequência a física moderna e contemporânea estará estabelecida em ambiente escolar, acompanhando o avanço científico. Foi mostrado também que a ideia de ondas gravitacionais em sala de aula, o que parece meio complexo e longe de suas realidades, se tornou algo compreensivo e de grande interesse, levantando questionamentos e trazendo de volta para o aluno aquela vontade de conhecer o funcionamento da física e o mundo que ela explica. Agostin~\cite{Agostin} menciona de uma forma bem geral tudo apresentado nessa seção em dizer que, 

\begin{quote}
(...) os alunos têm maturidade para aprender sobre essa temática e indicam que para que a mudança de fato ocorra em sala de aula é preciso dar condições e subsídios para que os professores tenham conhecimento sobre os assuntos, sintam-se confiantes para ensinar e possam assim contribuir para a renovação curricular de Física solicitada nos documentos oficiais. (pág. 5)
\end{quote}


\section{Conclusão}

Com base nos dados apresentados e discutidos observa-se a grande necessidade de renovação em que o currículo escolar do ensino de física precisa passar, sendo almejado não só por professores e pesquisadores nessa área, mas pelos próprios alunos que muitas vezes se sentem alienados por ficarem restrito a apenas um ensino ultrapassado, com didáticas ultrapassadas; alguns dos discente não se importam com tal ideal, justamente por já estarem desestimulados e desconhecidos de que o ensino de física atual continua a contribuir ainda mais para a sua completa rejeição dessa nova abordagem. Os próprios discentes sentem essa necessidade e veem na física moderna e contemporânea uma luz para superar essa barreira da dificuldade em física. O objetivo final desse trabalho foi mostrar que pessoas que não possuem nenhum contato com ciência de fronteira, possuem capacidade e anseio em aprender sobre coisas novas, conhecimentos revolucionários que trazem a sensação de que a física ainda está viva, pois infelizmente no ensino ela aparente está estagnada. O estereótipo inicial apresentado por esse trabalho deve ser quebrado. Cabe aos professores começarem a ter a audácia de desenvolver os alunos cientificamente, mostrando o outro lado da ciência e influenciando em suas escolhas futuras.

Ondas gravitacionais é um marco no século XXI, um marco de desenvolvimento científico e que precisa ser apresentado nas salas de aulas, pois como mencionado anteriormente, lá se encontra o maior polo de divulgação científica e precisa-se aproveitar tal oportunidade, não falando apenas em ondas gravitacionais, mas em assuntos como física de partículas, física nuclear, supercondutores, nanotecnologia e outros mais que devem ser incrementados ao ensino, trazendo nova luz ao conhecimento e seguindo o que já está previsto nas leis que regem a educação do país.

Todo e qualquer ser pensante possui capacidade de entender, compreender e aplicar evoluções e revolução na física, sendo que não só a comunidade científica deva se vangloriar com os avanços que estão a porta da sociedade e com os que virão, mas também o ensino deveria se alegrar e acompanhar tal momento em que a sociedade passa, no quesito desenvolvimento da física. Talvez essa seja a porta de entrada para melhorar o entendimento científico dos alunos e começar a não se preocupar mais com quantidade, mas sim dessa vez, com qualidade. Indícios para isso, esse trabalho mostra que há, precisando apenas por em prática tal idealização e começar a fazer a diferença no ensino de física brasileiro.

Desenvolver pesquisas físicas, acrescendo mais ainda ao nosso entendimento de Universo é de suma importância, mas resolver problemáticas da divulgação científica através do ensino aparenta ser uma problemática bem mais alarmante no Brasil, sendo que ainda, mesmo a longo prazo, pode trazer inúmeros frutos e avanços.


\section*{Agradecimentos}
Os autores agradecem ao Colégio São Lucas, Picos (PI), e ao Instituto Federal de Educação, Ciência e Tecnologia do Piauí - IFPI, \textit{Campus} Picos, pelo suporte e colaboração para a realização da pesquisa; e aos Professores Pedro José Feitosa Alves Júnior e Maria Girlandia de Sousa pela leitura crítica e sugestões para a melhoria da escrita do artigo.


\end{document}